\documentclass{emulateapj} 
\usepackage{apjfonts}
\usepackage{color}

\begin{document}

\title{Extreme particle acceleration in magnetic reconnection layers. Application to the gamma-ray flares in the Crab Nebula.}

\shorttitle{Extreme particle acceleration in magnetic reconnection layers}

\author{Beno\^it Cerutti$^1$, Dmitri~A.~Uzdensky$^1$, Mitchell~C.~Begelman$^{2,3}$} \shortauthors{Cerutti, Uzdensky, \& Begelman}

\affil{$^1$ CIPS, Physics Department, University of Colorado, UCB 390, Boulder, CO 80309-0390}
\email{benoit.cerutti@colorado.edu, uzdensky@colorado.edu, mitch@jila.colorado.edu}

\affil{$^2$ JILA, University of Colorado and National Institute of Standards and Technology, UCB 440, Boulder, CO 80309-0440}

\affil{$^3$ Department of Astrophysical and Planetary Sciences, University of Colorado, UCB 391, Boulder, CO 80309-0391}

\begin{abstract}
The gamma-ray space telescopes \emph{AGILE} and \emph{Fermi} detected short and bright synchrotron gamma-ray flares at photon energies above 100~MeV in the Crab Nebula. This discovery suggests that electron-positron pairs in the nebula are accelerated to PeV energies in a milliGauss magnetic field, which is difficult to explain with classical models of particle acceleration and pulsar wind nebulae. We investigate whether particle acceleration in a magnetic reconnection layer can account for the puzzling properties of the flares. We numerically integrate relativistic test-particle orbits in the vicinity of the layer, including the radiation reaction force, and using analytical expressions for the large-scale electromagnetic fields. As they get accelerated by the reconnection electric field, the particles are focused deep inside the current layer where the magnetic field is small. The electrons suffer less from synchrotron losses and are accelerated to extremely high energies. Population studies show that, at the end of the layer, the particle distribution piles up at the maximum energy given by the electric potential drop and is focused into a thin fan beam. Applying this model to the Crab Nebula, we find that the emerging synchrotron emission spectrum peaks above 100~MeV and is close to the spectral shape of a single electron. The flare inverse Compton emission is negligible and no detectable emission is expected at other wavelengths. This mechanism provides a plausible explanation for the gamma-ray flares in the Crab Nebula and could be at work in other astrophysical objects such as relativistic jets in active galactic nuclei.
\end{abstract}


\section{Introduction}\label{intro}

The conversion of magnetic to particle kinetic energy is a long standing problem in astrophysics, {\em e.g.} in relativistic jets and pulsar winds (the ``$\sigma$'' problem). Magnetic reconnection is the main mechanism known to dissipate magnetic energy into thermal and non-thermal energy of the plasma (see {\em e.g.} \citealt{2009ARA&A..47..291Z} for a recent review). A precise and accurate understanding of this process is still lacking, in particular for relativistic reconnection ({\em i.e.} if the Alfv\'en speed approaches the speed of light, $V_{\rm A}\approx c$), but current Particle-In-Cell (PIC) simulations have become a powerful tool to probe the microphysics of reconnection. Several of these PIC simulations were carried out in the context of relativistic electron-positron pair plasmas as found, for instance, in pulsar winds (see {\em e.g.} \citealt{2001ApJ...562L..63Z,2004ApJ...605L...9J,2007ApJ...670..702Z,2007A&A...473..683P,2008ApJ...682.1436L,2008ApJ...677..530Z,2011PhPl...18e2105L,2011ApJ...741...39S}). These studies demonstrated that magnetic reconnection is able to channel a sizeable fraction of the available magnetic energy into a population of non-thermal ultrarelativistic particles {\em via} the induced large-scale electric field.

The gamma-ray space telescopes \emph{AGILE} and \emph{Fermi} recently discovered bright day-long gamma-ray flares above 100~MeV in the Crab Nebula \citep{2011Sci...331..739A,2011Sci...331..736T,2011A&A...527L...4B,2011ApJ...741L...5S,2011arXiv1112.1979B}. This high-energy emission is thought to be synchrotron radiation by relativistic electron-positron pairs in the nebula. However, these flares exhibit several puzzling properties that challenge classical models of pulsar wind nebulae \citep{1974MNRAS.167....1R,1984ApJ...283..694K} and particle acceleration. One of the most intriguing features is the emission of synchrotron photons above the classical limit $\epsilon_{\star}=(9m_{\rm e} c^2/4\alpha_{\rm F})(E/B_{\perp})\approx 160~(E/B_{\perp})$~MeV (where $m_{\rm e}$ is the mass of the electron, $\alpha_{\rm F}$ is the fine structure constant, $E$ is the electric field and $B_{\perp}$ is the magnetic field perpendicular to the particle's motion) imposed by the balance between the accelerating electric force and the damping radiation-reaction force due to synchrotron energy losses \citep{1983MNRAS.205..593G,1996ApJ...457..253D}, assuming that $E<B_{\perp}$. These observations support the presence of PeV particles in the nebula (corresponding to Lorentz factor $\gamma_{\rm e}\sim 10^9$), the highest energy particles ever associated with a specific astrophysical source \citep{2011Sci...331..739A}. Another challenging aspect of the flares is that most of the observed gamma-ray emission above 100~MeV should originate from a tiny part of the nebula ($ct_{\rm flare}\sim 3\times 10^{15}$~cm~$\ll 0.1~$pc) if the emitting region is causally connected.

In a previous study \citep{2011ApJ...737L..40U}, we proposed that the gamma-ray flares in the Crab Nebula could be attributed to extreme particle acceleration in the vicinity of a reconnection layer present in the nebula. In this scenario, high-energy particles move across the reversing magnetic field and follow the relativistic analog of Speiser's trajectories \citep{1965JGR....70.4219S}. We found that the particles' orbits are naturally focused into the reconnection layer where the reconnecting magnetic field is small, in particular smaller than the reconnection electric field. Because of this, particles suffer weaker synchrotron losses and can be accelerated to extremely high energies deep inside the layer \citep{2004PhRvL..92r1101K}. Motivated by the results of our first study, here we examine in detail the process of particle acceleration in the electromagnetic field configuration likely to be found in a reconnecting system. We use relativistic test-particle simulations, {\em i.e.} the interaction between particles and the back-reaction of particles on the fields are neglected. Importantly, in the equation of motion we include the radiation reaction force induced by synchrotron and inverse Compton losses. The only other similar studies that also included the radiation reaction force are \citet{1998A&A...335...26S,1999PhPl....6.4318S}, and \citet{2003PhPl...10..835N} using resistive magnetohydrodynamics (MHD) simulations; and the PIC simulations by \citet{2009PhRvL.103g5002J}. In this paper, we focus specifically on the motion of the highest energy particles (with a Larmor radius greater than the thickness of the reconnection layer) present in the plasma and discuss the possible application of this process to the gamma-ray flares in the Crab Nebula.

This article is organized as follows. In Section~\ref{reconnection}, the main assumptions and approximations adopted for the reconnection layer in a relativistic pair plasma are presented. Section~\ref{rad_force} is dedicated to the equations of motion of relativistic particles in the reconnection layer, with a particular emphasis on the expression for the radiation reaction force. Section~\ref{single_electron} focuses on the motion of a single particle and analyzes the effect of each large-scale electromagnetic field component, and synchrotron and Compton losses. We study the distribution of a population of electrons at the end of the reconnection layer in Section~\ref{pop}, and apply the model to the Crab Nebula in Section~\ref{crab}. The final part (Section~\ref{ccl}) summarizes the results of this study.

\section{The relativistic reconnection layer}\label{reconnection}

\begin{figure}
\epsscale{1.0}
\plotone{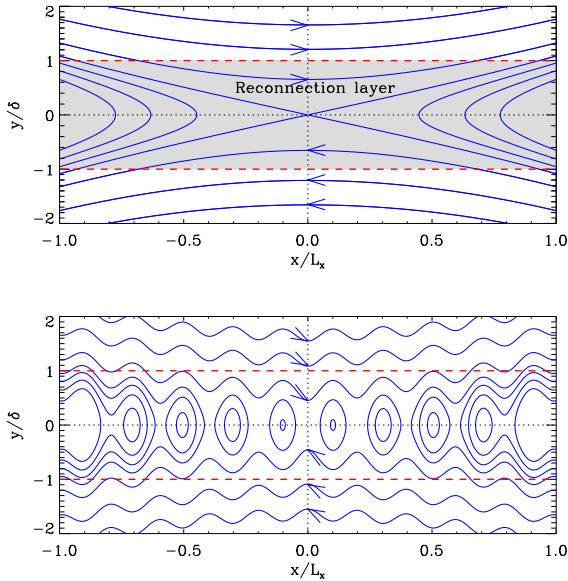}
\caption{\emph{Top:} Diagram of the magnetic field line configuration (isocontour of the magnetic flux function $\psi$) at the reconnection site in the $xy$-plane containing the reconnecting ($B_{\rm x}$) and the reconnected ($B_{\rm y}$) components of the magnetic field. The reconnection layer is delimited by the grey region between $y=\pm\delta$. {\em Bottom:} Same as in the {\em top} panel, but where the reconnection layer is broken up into a chain of magnetic islands.}
\label{fig_fields}
\end{figure}

\subsection{Emphasis on the highest energy particles}

In the present study, we are interested in the dynamics of the highest energy particles present in the reconnection region, namely, the particles with a Larmor radius much larger than the thickness of the layer. These particles are assumed to be preaccelerated by some unspecified mechanism. In the context of the Crab Nebula gamma-ray flares, the emitting particles have a typical Lorentz factor $\gamma_{\rm e}\gtrsim 10^9$ and a gyroradius $\rho_{\rm e}\sim 10^{15}~$cm (in a milli-Gauss field), of order the size of the emitting region $L\sim ct_{\rm flare}\sim 3\times 10^{15}~$cm that we associate here with the length of the reconnection layer $L_{\rm z}$. The emphasis on such particles is unique, because one does not usually face the situation where $\rho_{\rm e}\sim L_{\rm z}$ in other astrophysical environments, {\em e.g.} in the solar corona. Because of their large Larmor radii, the particles of interest here feel only the large-scale field structure as described below. Small-scale ({\em e.g.} turbulent) structures in the current layer are probably important for acceleration of lower energy particles, but do not strongly affect the motion of the highest-energy particles. This justifies the assumptions of using only the large-scale fields in this study, and significantly simplifies the problem.

\subsection{The large-scale electromagnetic fields}

We consider a simple geometry of a steady state and laminar magnetic reconnection region in a collisionless plasma of relativistic pairs (Fig.~\ref{fig_fields}). The large-scale electromagnetic field consists of the reconnecting magnetic field $B_{\rm x}$ in the $x$-direction, the reconnection electric field $E_{\rm z}$ in the $z$-direction, and the reconnected magnetic field $B_{\rm y}$ in the $y$-direction. In the general case, there is also a component of the magnetic field along the electric field $B_{\rm z}$, called the guide field, and the associated in-plane induced electric field. The guide field $B_{\rm z}$ has a major effect in suppressing relativistic drift-kink instabilities in the reconnection layer and thus may promote non-thermal particle acceleration \citep{2008ApJ...677..530Z}. Because there is no Hall effect in pair plasmas, the quadrupole out-of-plane Hall magnetic field and the associated electrostatic in-plane electric field components (important in electron-ion reconnection) are absent.

The current layer has a slab-like geometry with a characteristic thickness $\delta$ in the $y$-direction across which the magnetic field lines reverse and reconnect. In collisionless pair reconnection, the lower limit on the thickness of the layer is set by the relativistic electron skin depth $d_{\rm e}=(\gamma_{\rm av} m_{\rm e} c^2/4\pi n_{\rm e} e^2)^{1/2}$ (where $n_{\rm e}$ is the electron density and $\gamma_{\rm av}$ is the characteristic Lorentz factor of the particles in the plasma) if there is no strong guide field. Otherwise, $\delta$ is given by the Larmor radius of the bulk electrons $\delta=\gamma_{\rm av}\rho_0$, where $\rho_0=m_{\rm e} c^2/e B_0$. These two scales are comparable if the guide field strength is similar to (or smaller than) the reconnecting magnetic field and if the upstream plasma pressure is not large compared with the magnetic pressure. Note that if the layer is very long (compared with its thickness $\delta$), it is subject to instabilities ({\em e.g.} tearing) that may lead to its effective broadening. In this case, and assuming that the plasma is incompressible, the effective thickness of the layer is at most $\delta=\beta_{\rm rec}L_{\rm x}$ ({\em e.g.} \citealt{2001EP&S...53..473S}), where $\beta_{\rm rec}<1$ is the dimensionless reconnection rate and $L_{\rm x}$ is the size of the layer in the $x$-direction.

The reconnecting magnetic field $B_{\rm x}$ is roughly uniform outside the reconnection layer ($|y|\gg \delta$) and equals $\pm B_0$. Inside the layer, the amplitude of $B_{\rm x}$ decreases and by symmetry vanishes at $y=0$. The field is commonly modeled by a Harris profile,
\begin{equation}
B_{\rm x}\left(y\right)=B_0 \tanh\left(\frac{y}{\delta}\right).
\label{bx}
\end{equation}
The electric field induced by magnetic reconnection is roughly uniform in the reconnection region in a steady state, and is a finite fraction of the reconnecting magnetic field: $E_{\rm z}=\beta_{\rm rec}B_0$. The reconnection rate $\beta_{\rm rec}$ is poorly constrained in the context of collisionless relativistic pair plasmas but a few simulations estimate $\beta_{\rm rec}\sim 0.1$ ({\em e.g.} \citealt{2008ApJ...677..530Z}). We will adopt this value in this paper.

\begin{figure}
\epsscale{1.0}
\plotone{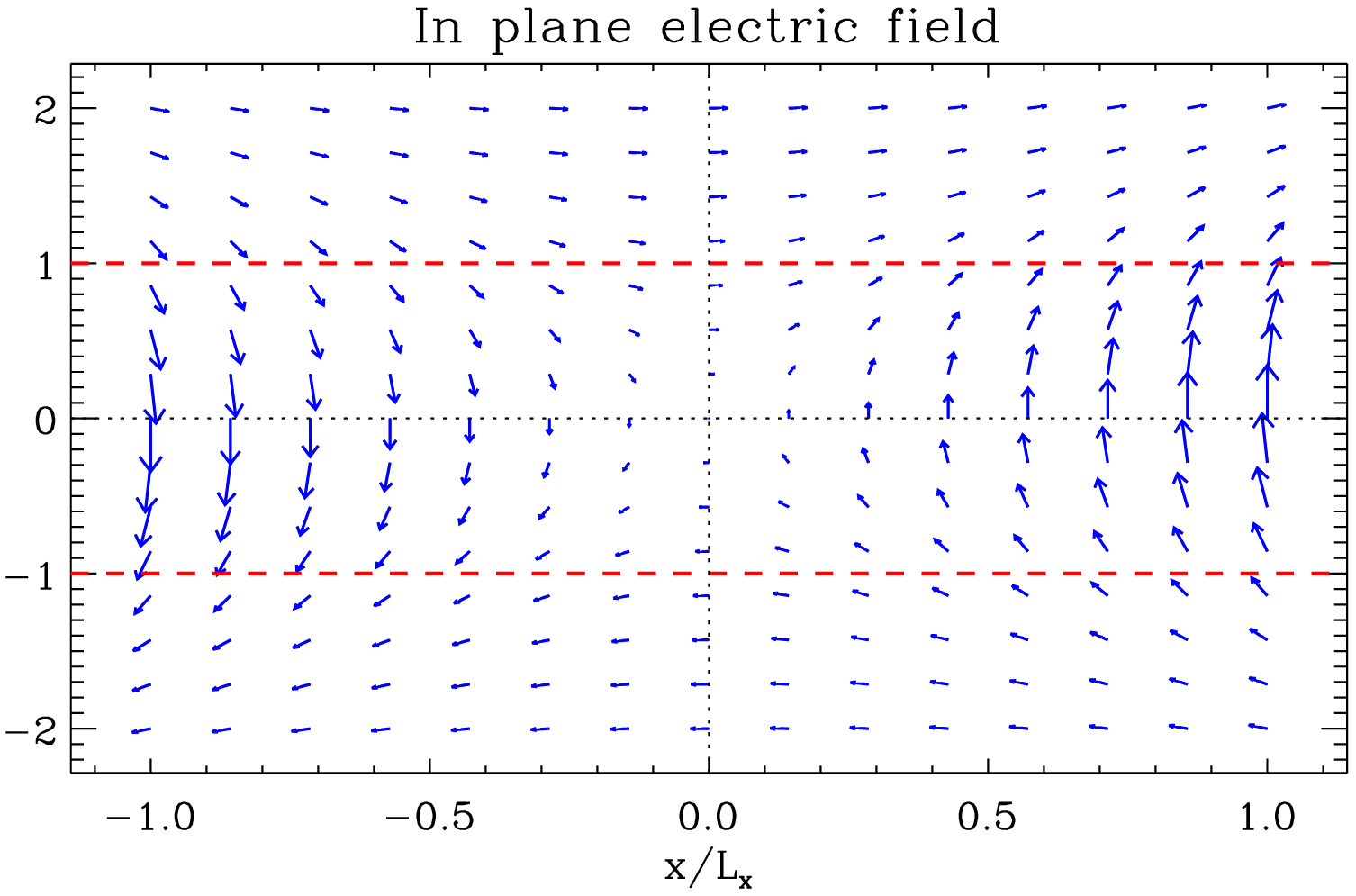}
\caption{The blue arrows represent the electric field vector in the $xy$-plane induced by the motion of the plasma through the reconnection layer in the presence of a guide field.}
\label{fig_electric}
\end{figure}

In the presence of the guide field, the motion of the plasma in the reconnection layer generates an electric field in the $xy$-plane \citep{1999PhPl....6.4318S}, such that $\mathbf{E}=-\mathbf{V}\times \mathbf{B_{\rm z}}/c$. Here, $\mathbf{V}$ is the velocity of the plasma in the $xy$-plane which we model as (assuming that the Alfv\'en speed $V_{\rm A}\approx c$)
\begin{eqnarray}
V_{\rm x} &=& c\left(\frac{x}{L_{\rm x}}\right)\cosh^{-2}\left(\frac{y}{\delta}\right) \\
V_{\rm y} &=& -c\beta_{\rm rec}\tanh\left(\frac{y}{\delta}\right).
\end{eqnarray}
These components of the electric field $E_{\rm x}$ and $E_{\rm y}$ are included in this study (see Fig.~\ref{fig_electric} for a visual representation of the electric field structure in the $xy$-plane). Any bulk motion of the plasma in the $z$-direction is neglected, $V_{\rm z}=0$.

The structure of the reconnected magnetic field $B_{\rm y}$ is much more uncertain than that of the other large-scale fields. The reconnection region can be viewed as a long Sweet-Parker-like layer of length $L_{\rm z}\gg \delta$ and width along the $x$-direction $2L_{\rm x}$, which is generally similar to $L_{\rm z}$. In the layer ($\left|x\right|<L_{\rm x}$), the large-scale reconnected magnetic field behaves approximately as
\begin{equation}
B_{\rm y}\left(x\right)=\beta_{\rm rec} B_0\left(\frac{x}{L_{\rm x}}\right),
\label{by}
\end{equation}
reaching $\beta_{\rm rec} B_0$ at the end of the layer, $|x|=L_{\rm x}$. Outside the reconnection layer ($|x|>L_{\rm x}$), $B_{\rm y}$ increases rapidly with $|x|$ to reach $B_0$ over a distance comparable to $L_{\rm x}$.

\subsection{Tearing and kink instabilities in the layer}

This simple picture may be a reasonable approximation for the average large-scale field structure in the reconnection region, but it may not capture its small and intermediate scale structure. In the Sweet-Parker picture \citep{1958IAUS....6..123S,1957JGR....62..509P}, the expected reconnection rate would be very small, much smaller than 0.1. Petschek-like reconnection \citep{1964NASSP..50..425P} would be much faster but this configuration is unlikely to appear in pair plasmas because there is no Hall effect (see, however, \citealt{2005PhRvL..95x5001B,2007PhPl...14e6503B}). Analytical calculations and numerical simulations show instead that a thin Sweet-Parker-like layer is unstable to secondary tearing modes and breaks up into a highly dynamical chain of magnetic islands (or plasmoids) connected by small current layers (see {\em e.g.}, \citealt{2007PhPl...14j0703L,2009PhRvL.103j5004S,2009PhRvL.103f5004D}). In this regime, magnetic reconnection is fast (see {\em e.g.}, \citealt{2001EP&S...53..473S,2009PhPl...16k2102B,2009PhPl...16l0702C,2009PhRvL.103f5004D,2010PhPl...17f2104H,2010PhRvL.105w5002U,2011arXiv1108.4040L}) and $\beta_{\rm rec}$ could be 0.1 \citep{2008ApJ...677..530Z}. It is important to note that each magnetic island contributes to $B_{\rm y}$, but the net reconnected magnetic flux within an island is zero because the magnetic field lines are closed loops (Fig.~\ref{fig_fields}, {\em bottom} panel). The effective large-scale reconnecting magnetic field would result from the semi-open reconnected field lines generated between the current layers and the islands \citep{2010PhRvL.105w5002U}.

A simple way to quantify the effect of magnetic islands on the propagation of particles is to add a static perturbation $\tilde{\psi}$ to the undisturbed magnetic flux function $\psi_0$ (this quantity corresponds to the $z$-component of the magnetic vector potential $A_{\rm z}$) of the form
\begin{equation}
\tilde{\psi}=\epsilon B_0\delta\cos\left[n_{\rm is}\pi\left(\frac{x}{L_{\rm x}}\right)\right],
\end{equation}
where $n_{\rm is}$ is the number of islands and $\epsilon$ is the amplitude of the perturbation normalized to $B_0 \delta$, both quantities considered here as free parameters. Only the $y$-component of the magnetic field is perturbed, and equals
\begin{equation}
\tilde{B_{\rm y}}=-\frac{\partial\tilde\psi}{\partial x} = \epsilon n_{\rm is} \pi B_0 \left(\frac{\delta}{L_{\rm x}}\right)\sin\left[n_{\rm is}\pi\left(\frac{x}{L_{\rm x}}\right)\right].
\label{by_is}
\end{equation}
The overall smoothed growth of the reconnected magnetic field would not be very different from a linear variation across the layer as described by Eq.~(\ref{by}).

A thin current layer can also bend along the $z$-direction due to the relativistic drift kink instability ({\em e.g.} \citealt{2005ApJ...618L.111Z,2011PhPl...18e2105L}), unless there is a strong guide field. We describe the effect of the kink instability as a stationary sinusoidal translation of the current layer in the $y$-direction as a function of $z$, so that the reconnecting magnetic field is given by
\begin{equation}
B_{\rm x}\left(x,y\right)=B_0\tanh\left[\frac{y}{\delta}-C_{\rm y}\sin\left(\frac{2\pi n_{\rm z} z}{L_{\rm z}}\right)\right],
\label{eq_kink}
\end{equation}
where $C_{\rm y}$ is the amplitude of the deformation of the layer in the $y$-direction and $n_{\rm z}$ is the number of oscillations along the layer. This simple prescription is a rather good representation of the layer in the linear stage of the kink instability. The other electromagnetic fields are unchanged.

\section{Radiation reaction force and equations of motion}\label{rad_force}

This section gives the complete expressions for the equations of motion of a single electron moving in an arbitrary electromagnetic field and subject to radiative damping (Section~\ref{sect_motion}). Sections~\ref{sect_gen}-\ref{sect_syn} focus on general and approximate expressions of the radiation reaction force. Section~\ref{sect_drag} provides the expression of the inverse Compton drag force in the Thomson regime.

\subsection{General expression for the radiation reaction force}\label{sect_gen}

Independently of the underlying acceleration mechanism, a relativistic particle of charge $e$ and Lorentz factor $\gamma_{\rm e}$ loses energy at a rate given by the Larmor formula (in covariant form, see \emph{e.g.} \citealt{1975clel.book.....J})
\begin{equation}
\frac{{\rm d}\mathcal{E}_{\rm rad}}{{\rm d}t}=\mathcal{P}_{\rm rad}=\frac{2}{3} e^2 c\left(\frac{{\rm d} u^{\mu}}{{\rm d}s}\right)\left(\frac{{\rm d} u_{\mu}}{{\rm d}s}\right),
\label{eq_losses}
\end{equation}
where $u^{\mu}=(\gamma_{\rm e},\gamma_{\rm e}\mathbf{v}/c)$ is the four-velocity, and the interval ${\rm d}s=c {\rm d}t/\gamma_{\rm e}$ ($t$ is the time in the laboratory frame). This expression can be reformulated in terms of the components of the three-acceleration $\mathbf{a}$ parallel $(a_{\parallel})$ and perpendicular $(a_{\perp})$ to the direction of motion of the particle as \citep{1979rpa..book.....R}
\begin{equation}
\mathcal{P}_{\rm rad}=\frac{2e^2}{3c^3} \gamma_{\rm e}^4\left(a_{\perp}^2+\gamma_{\rm e}^2 a_{\parallel}^2\right).
\label{eq_losses2}
\end{equation}
In return, radiative energy losses affect the dynamics of the particle through the radiative back-reaction four-force $g^{\mu}=(\gamma_{\rm e}\mathbf{g}\cdot\mathbf{v}/c^2,\gamma_{\rm e}\mathbf{g}/c)$ given by \citep{1975clel.book.....J}
\begin{equation}
g^{\mu}=\frac{2 e^2}{3c}\frac{{\rm d}^2 u^{\mu}}{{\rm d}s^2}-\frac{\mathcal{P}_{\rm rad}}{c^2}u^{\mu}.
\label{gmu}
\end{equation}
The first term in this equation is known as the Schott term and the second term describes the recoil of the particle due to radiative losses. In the ultrarelativistic limit ($\gamma_{\rm e}\gg 1$, the regime we are interested in here), the Schott term is negligible and the spatial component of $g^{\mu}$ can be expressed as \citep{1971ctf..book.....L}
\begin{equation}
\mathbf{g}\approx-\frac{\mathcal{P}_{\rm rad}}{c^2}\mathbf{v},
\label{gmu_approx}
\end{equation}
and acts as a friction force opposite to the direction of motion of the particle. If the radiative losses term vanishes, $\mathcal{P}_{\rm rad}=0,$ the Schott term determines the strength of the radiation reaction force. But then, the radiation reaction force is negligible compared with the electric force in the equation of motion (see below).

\subsection{Radiation reaction force in an electromagnetic field}\label{sect_syn}

If one now considers the specific situation of a relativistic electron of mass $m_{\rm e}$ moving in a given electromagnetic field $(\mathbf{E},\mathbf{B})$, the equation of motion is (Lorentz-Abraham-Dirac equation in the laboratory frame, \citealt{1975clel.book.....J})
\begin{equation}
m_{\rm e} c \frac{{\rm d} u^{\mu}}{{\rm d}s}=-\frac{e}{c}F^{\mu\nu}u_{\nu}+g^{\mu},
\label{eq_lorentz}
\end{equation}
where $F^{\mu\nu}$ is the external electromagnetic field-strength tensor. In the framework of classical electrodynamics, \citet{1971ctf..book.....L} derived an approximate expression for $\mathbf{g}$ in terms of the external electromagnetic fields using an iterative procedure. First, the radiation reaction four-force is neglected in the equation of motion, $g^{\mu}=0$ and only the acceleration due to the Lorentz force contributes to the radiative energy losses. Substituting the four-acceleration term ${\rm d}u^{\mu}/{\rm d}s$ by $-(e/m_{\rm e}c^2)F^{\mu\nu}u_{\nu}$ in Eq.~(\ref{eq_losses}) yields
\begin{equation}
\mathcal{P}_{\rm rad}=\frac{2}{3}r_{\rm e}^2 c \gamma_{\rm e}^2\left[\left(\mathbf{E}+\frac{\mathbf{v}\times\mathbf{B}}{c}\right)^2-\left(\frac{\mathbf{v}\cdot\mathbf{E}}{c}\right)^2\right],
\label{eq_losses_em}
\end{equation}
where $r_{\rm e}=e^2/m_{\rm e} c^2$ is the classical radius of the electron. The term proportional to $B^2$ in Eq.~(\ref{eq_losses_em}) is the classical synchrotron energy loss related to the perpendicular acceleration by the magnetic field ($a_{\perp}\approx eB_{\perp}/\gamma_{\rm e}m_{\rm e}$, $B_{\perp}$ is the component of the magnetic field perpendicular to the particle's direction of motion). This is the dominant term in most cases. The other terms are associated with the linear acceleration by the electric field and the perpendicular acceleration by the electric field and the magnetic field. These terms are included in our calculations although we find that they do not lead to significant changes.

In the first order approximation for $\gamma_{\rm e}\gg 1$, the radiation reaction force is
\begin{equation}
\mathbf{g}=-\frac{\mathcal{P}_{\rm rad}}{c^2}\mathbf{v}.
\label{g3}
\end{equation}
The higher order terms in the expression of the force can be obtained by successive iterations (by injecting into Eq.~(\ref{eq_losses}) the Lorentz force and the first order radiation reaction force term). In the second order (in $\gamma_{\rm e}^{-2}$), the radiation reaction force contributes to the linear acceleration term in the Larmor formula (Eq.~\ref{eq_losses2}) such that $\gamma_{\rm e}a_{\parallel,\rm{rad}}=|\mathbf{g}|/\gamma_{\rm e}^2 m_{\rm e}$ ($a_{\perp}$ is unchanged). Hence, the first order formula in Eq.~(\ref{g3}) is valid as long as $a_{\perp}\gg\gamma_{\rm e}a_{\parallel,\rm{rad}}$, implying that
\begin{equation}
\frac{\gamma_{\rm e}B}{B_{\rm c}}\ll 1,
\label{b}
\end{equation}
{\em i.e.}, if the magnetic field in the rest frame of the particle does not exceed the classical critical magnetic field strength $B_{\rm c}=m^2_{e}c^4/e^3\approx 6\times 10^{15}~$G (above which the formal Larmor radius of the particle in the rest frame $\rho_{\rm e}=m_{\rm e}c^2/eB$ becomes smaller than $r_{\rm e}$). In reality, quantum effects become important at a much smaller field $B_{\rm QED}=m^2_{\rm e}c^3/\hbar e=\alpha_{\rm F}B_{\rm c}\approx 4.4\times 10^{13}~$G (at which $\rho_{\rm e}$ is comparable to the reduced Compton wavelength $\lambdabar_{\rm e}=\hbar/m_{\rm e} c$) and so the condition in Eq.~(\ref{b}) should be reformulated as $b\equiv\gamma_{\rm e}B/B_{\rm QED}\ll 1$. In the quantum regime ($b\gg 1$), the electron emits in discrete steps photons with an energy comparable to the electron's energy \citep{1972PhRvD...6.2736S} and the radiation reaction force should be modeled as a stochastic process where the electron undergoes discrete and substantial energy losses $\Delta\gamma_{\rm e}/\gamma_e\approx 1$ (see {\em e.g.} \citealt{2011PPCF...53a5009D}). This situation would require a full quantum electrodynamics treatment which is not undertaken in this article.

It is worthwhile to note that the radiation reaction three-force can be comparable or even exceed the Lorentz three-force $\mathbf{F_{\rm L}}$ without violating the condition $b\ll 1$. Indeed, with $|\mathbf{g}|\sim r_{\rm e}^2\gamma_{\rm e}^2 B^2$ and $|\mathbf{F_{\rm L}}|\sim e B$, the ratio of the two forces scales as
\begin{equation}
\frac{|\mathbf{g}|}{|\mathbf{F_{\rm L}}|}\sim \frac{\gamma_{\rm e}^2 B}{B_{\rm c}}=\alpha_{\rm F}\gamma_{\rm e}b.
\label{ratio_forces}
\end{equation}
Hence, if $|\mathbf{g}|\sim|\mathbf{F_{\rm L}}|$ and $b\ll 1$, then the electron has to be ultrarelativistic with a Lorentz factor $\gamma_{\rm e}>\alpha^{-1}_{\rm F}\gg 1$. In the context of the gamma-ray flares in the Crab Nebula, where PeV electrons ($\gamma_{\rm e}\sim 10^9$) evolve in a milli-Gauss magnetic field (see Section~\ref{crab}), $b\sim 10^{-7}$ and we are well within the range of validity of Eq.~(\ref{g3}) even when the particles reach the radiation reaction limit regime where $|\mathbf{g}|\sim |\mathbf{F_{\rm L}}|$.

\subsection{Inverse Compton drag force in the Thomson regime}\label{sect_drag}

If the electron is bathed in an external radiation field, it loses energy {\em via} inverse Compton scattering. In the Thomson regime ($\epsilon_0^{'}/m_{\rm e} c^2\ll 1$, where $\epsilon_0^{'}$ is the target photon energy in the rest frame of the electron), the energy loss in each scattering event is small, $\Delta\gamma_{\rm e}/\gamma_{\rm e}\ll 1$, and the Compton drag force can be described as a continuous force. If the photon field is isotropic in the laboratory frame, then the Compton power radiated by a single electron is \citep{1970RvMP...42..237B}
\begin{equation}
\mathcal{P}_{\rm ic}=\frac{4}{3}\sigma_{\rm T} c \gamma_{\rm e}^2 \mathcal{U}_{\star},
\label{eq_losses_ic}
\end{equation}
where $\sigma_{\rm T}=(8/3)\pi r_e^2$ is the Thomson cross section and $\mathcal{U_{\star}}$ is the energy density of the radiation field (in erg~$\rm{cm}^{-3}$). In the rest frame of the particle, the target photons are focused in a cone of semi-aperture angle $\sim 1/\gamma_{\rm e}$ in the direction opposite to the motion of the particle. Hence, the Compton drag force can be formally written as a friction force such that
\begin{equation}
\mathbf{f}_{\rm ic}=-\frac{\mathcal{P}_{\rm ic}}{c^2}\mathbf{v}.
\label{drag_ic}
\end{equation}
The continuous drag force approximation is not valid in the Klein-Nishina regime ($\epsilon_0^{'}/m_{\rm e} c^2\gg 1$) in which the electron loses a significant fraction of its energy in a single interaction (as for synchrotron losses in the quantum regime).

\subsection{Equations of motion}\label{sect_motion}

Summarizing the discussion in the preceding sections, the equations describing the time evolution of an ultrarelativistic electron are (in the laboratory frame, and for $b\ll 1$)
\begin{eqnarray}
\label{eq_motion}
\frac{{\rm d}\mathbf{p}}{{\rm d}t} & = & -e\left(\mathbf{E}+\frac{\mathbf{v\times\mathbf{B}}}{c}\right)-\left(\frac{\mathcal{P_{\rm rad}}+\mathcal{P_{\rm ic}}}{c^2}\right)\mathbf{v} \\
\frac{{\rm d}\mathcal{E}_{\rm e}}{{\rm d}t} & = & -e\left(\mathbf{v}\cdot\mathbf{E}\right)-\mathcal{P}_{\rm rad}-\mathcal{P}_{\rm ic}.
\label{eq_energy}
\end{eqnarray}
In these equations $\mathbf{p}=\gamma_{\rm e} m_{\rm e} \mathbf{v}$ is the momentum and $\mathcal{E}_{\rm e}=\gamma_{\rm e} m_{\rm e} c^2$ is the total energy of the electron. In the following, we will refer to Eqs.~(\ref{eq_motion}-\ref{eq_energy}) as the equations of motion.

\section{Motion of a single electron in the magnetic reconnection configuration}\label{single_electron}

We solve the equations of motion including the radiation reaction in the vicinity of the reconnection layer described in Section~\ref{reconnection}. We examine in detail the effect of each component of the magnetic field on the dynamics of a single particle, starting with the reconnecting component $B_{\rm x}$ only. We first review the main solutions of the analytic model of \citet{2011ApJ...737L..40U} (Section~\ref{analytical}). Analytical solutions are then compared with the numerical solutions (Section~\ref{compare}). We then investigate the effects of the guide field components (Section~\ref{sect_bz}), the reconnected field and the presence of magnetic islands in the current sheet (Section~\ref{sect_by}), and kinking of the current sheet (Section~\ref{sect_kink}) on the motion of the particle. The effect of Compton scattering is discussed in Section~\ref{effect-compton}.

\subsection{Analytic solutions in the reconnecting fields only}\label{analytical}

\citet{2011ApJ...737L..40U} studied the evolution of a single electron, assuming that the guide field and the reconnected magnetic field are negligible (and without magnetic islands), and neglecting the inverse Compton drag force. In this model, the electron is initially injected at the origin of the axes with a Lorentz factor $\gamma_{0}\gg 1$ and a velocity contained in the $yz$-plane crossing the $z$-direction with an angle $\theta_{0}<\pi/2$ such that $v_{\rm y}(z=0)\approx c\sin\theta_0$ (see Fig.~\ref{fig_rec}). If the Larmor radius of the electron is much greater than the thickness of the layer, the trajectory of the electron is composed of portions of cyclotron orbits stretched along the $z$-direction by the reconnection electric field. The reversing reconnecting magnetic field always deflects the particle towards the midplane $y=0$. As the electron gains energy, the Larmor radius of the particle increases but the maximum deviation of the orbit from the midplane $y_{\rm max}$ decreases.

It is possible to derive simple analytical solutions to the equations of motion Eqs.~(\ref{eq_motion}-\ref{eq_energy}) in three special regimes:
\begin{figure}
\epsscale{1.0}
\plotone{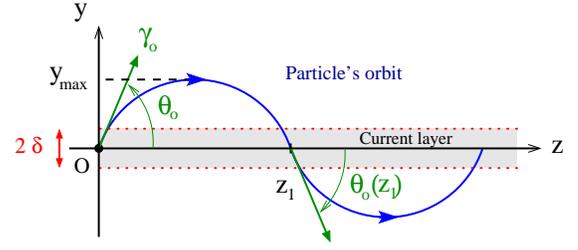}
\caption{Geometrical quantities defining the orbit of a single electron in the vicinity of the reconnection layer (relativistic Speiser orbit) in the $yz$-plane.}
\label{fig_rec}
\end{figure}
\newline
\newline
{\bf (1)} Most of the particle's trajectory is well outside the reconnection layer ($y_{\rm max}\gg \delta$) and its energy is well below the radiation reaction limit ($e c E_0\gg \mathcal{P}_{\rm rad}$), so that radiative losses can be ignored. Taking the $y$-component of Eq.~(\ref{eq_motion}) and writing ${\rm d}/{\rm d}t=v_{\rm z}{\rm d}/{\rm d}z$ and defining $\omega_0=eB_0/m_{\rm e} c$, we obtain (over one half-cycle, where $y>0$)
\begin{equation}
\frac{{\rm d}}{{\rm d}z}\left(\gamma_{\rm e}v_{\rm y}\right)=-\omega_0.
\label{reg1_1}
\end{equation}
Integrating this equation yields
\begin{equation}
\gamma_{\rm e}\beta_{\rm y}=-\bar{z}+\gamma_0\beta_{\rm y,0},
\label{reg1_2}
\end{equation}
where $\bar{z}=z/\rho_0$, $\beta_{\rm y}=v_{\rm y}/c$ and $\beta_{\rm y,0}=\beta_{\rm y}(0)\approx\sin\theta_0$. Considering, for simplicity, the case $\theta_0\ll 1$, so that $\beta_{\rm z}=v_{\rm z}/c\approx 1$ and $\beta_{\rm y}\approx {\rm d}\bar{y}/{\rm d}\bar{z}$ (with $\bar{y}=y/\rho_0$), Eq.~(\ref{reg1_2}) becomes
\begin{equation}
\gamma_{\rm e}\frac{{\rm d}\bar{y}}{{\rm d}\bar{z}}=-\bar{z}+\gamma_0\theta_0.
\label{reg1_3}
\end{equation}
The evolution of the Lorentz factor of the particle follows (after integration of Eq.~(\ref{eq_energy}))
\begin{equation}
\gamma_{\rm e}\left(\bar{z}\right)=\beta_{\rm rec}\bar{z}+\gamma_0.
\label{reg1_4}
\end{equation}
Combining Eq.~(\ref{reg1_3}) and Eq.~(\ref{reg1_4}), the final integration over $\bar{z}$ gives the orbit of the electron over one half cycle
\begin{equation}
\bar{y}\left(\bar{z}\right)=-\frac{\bar{z}}{\beta_{\rm rec}}+\left(\frac{\gamma_0}{\beta^2_{\rm rec}}\right)\left(1+\beta_{\rm rec}\theta_0\right)\ln\left(1+\frac{\beta_{\rm rec}\bar{z}}{\gamma_0}\right).
\label{reg1_5}
\end{equation}
If the amount of energy gained by the particle over one half cycle is small such that $\Delta\gamma_{\rm e}/\gamma_0=\beta_{\rm rec} \bar{z}/\gamma_0\ll 1$, then the trajectory of the electron follows (after an expansion of Eq.~(\ref{reg1_5}) to third order in $\beta_{\rm rec} \bar{z}/\gamma_0$)
\begin{equation}
\bar{y}\left(\bar{z}\right)\approx\theta_0\bar{z}-\frac{\bar{z}^2}{2\gamma_0}\left(1+\beta_{\rm rec}\theta_0\right)+\frac{1}{3}\frac{\beta_{\rm rec}\bar{z}^3}{\gamma_0^2}.
\label{reg1_6}
\end{equation}
At the end of the half cycle, the particle crosses the $z$-axis at a distance approximately equal to
\begin{equation}
\bar{z}_1\approx 2\gamma_0\theta_0+\frac{2}{3}\gamma_0\beta_{\rm rec}\theta^2_0.
\label{reg1_7}
\end{equation}
The variation of the Lorentz factor and the midplane crossing angle over one half cycle are then
\begin{eqnarray}
\Delta\gamma_{\rm e}&=&\gamma_{\rm e}\left(\bar{z}_1\right)-\gamma_0\approx 2\beta_{\rm rec}\gamma_0\theta_0 \\
\Delta\left|\theta_0\right|&=&\left|\theta_0\left(\bar{z}_1\right)\right|-\theta_0\approx -\frac{4}{3}\beta_{\rm rec}\theta_0^2.
\end{eqnarray}
The evolution of the crossing angle $\theta_0$ over many cycles ($\bar{z}\gg\bar{z}_1$) is governed by ${\rm d}|\theta_0|/{\rm d}\bar{z}\approx\Delta\left|\theta_0\right|/\bar{z}_1$ and hence after integration one finds that $\left|\theta_0\right|\propto \left(\gamma_{\rm e}\right)^{-2/3}\propto \left(\bar{z}\right)^{-2/3}$. A similar derivation gives that the maximum distance from the midplane decreases as $y_{\rm max}\propto \left(\gamma_{\rm e}\right)^{-1/3}\propto\left(\bar{z}\right)^{-1/3}$. In this regime, the energy of the particle increases and the particle's orbit shrinks steadily towards the midplane.
\newline
\newline
{\bf (2)} The particle reaches the radiation-reaction limit ($e c E_0= \mathcal{P}_{\rm rad}$) while well outside the layer ($y_{\rm max}\gg \delta$). This means that the energy gained by the electric acceleration is completely radiated away by the electron in each cycle. The Lorentz factor of the electron over each cycle is constant and is determined by the balance between the acceleration rate and the synchrotron energy loss (neglecting the energy loss by the electric field in Eq.~(\ref{eq_losses_em})), {\em i.e.}
\begin{equation}
\gamma_{\rm e}\equiv\gamma_{\rm rad}=\left(\frac{3c\beta_{\rm rec}}{2 r_{\rm e} \omega_{0}}\right)^{1/2}.
\label{grad}
\end{equation}
This is the standard radiation reaction limit, resulting in the synchrotron photon energy limit at $\sim 160~(E_0/B_0)~$MeV introduced in Section~\ref{intro}. However, in spite of the constant energy of the electron, the orbit continues to shrink towards the midplane because of the radiation reaction force. Using the same notation as in regime (1) and replacing $\gamma_{\rm e}$ with the expression for $\gamma_{\rm rad}$, the $y$-component of the equation of motion becomes
\begin{equation}
\gamma_{\rm rad}\frac{{\rm d}\beta_{\rm y}}{{\rm d}\bar{z}}=-1-\beta_{\rm rec}\beta_{\rm y}.
\end{equation}
After integration (at $t=0$, $\beta_{\rm y}(0)\approx\sin\theta_0\approx \theta_0$, assuming $\theta_0\ll 1$), we have
\begin{equation}
1+\beta_{\rm rec}\beta_{\rm y}\left(\bar{z}\right)=(1+\beta_{\rm rec}\theta_0)\exp\left(-\frac{\beta_{\rm rec}\bar{z}}{\gamma_{\rm rad}}\right).
\end{equation}
Expanding this solution to the second order in $\beta_{\rm rec}\bar{z}/\gamma_{\rm rad}\ll 1$, and integrating over $\bar{z}$, the equation of the trajectory becomes
\begin{equation}
\bar{y}\left(\bar{z}\right)\approx \theta_0\bar{z}-\frac{\bar{z}^2}{2\gamma_0}\left(1+\beta_{\rm rec}\theta_0\right)+\frac{1}{6}\frac{\beta_{\rm rec}\bar{z}^3}{\gamma_{\rm rad}^2}.
\end{equation}
The trajectory intersects the $y=0$ plane at a distance
\begin{equation}
\bar{z}_1\approx2\gamma_{\rm rad}\theta_0-\frac{2}{3}\gamma_{\rm rad}\beta_{\rm rec}\theta_0^2.
\end{equation}
Over one half cycle, the midplane crossing angle varies by
\begin{equation}
\Delta\left|\theta_0\right|=-\frac{2}{3}\beta_{\rm rec}\theta_0^2.
\end{equation}
The long term evolution of $\theta_0$ over many cycles follows ${\rm d}\theta_0/{\rm d}\bar{z}\approx\Delta\left|\theta_0\right|/\bar{z}_1$, hence the midplane crossing angle decreases exponentially as $\theta_0\propto \exp\left(-\beta_{\rm rec}\bar{z}/3\gamma_{\rm rad}\right)$.
\newline
\newline
{\bf (3)} In the third case, one considers an electron whose orbit is entirely confined deep inside the layer ($y_{\rm max}\ll \delta$) where the magnetic field varies approximately linearly $B_{\rm x}(y)\approx(y/\delta)B_0$. The maximum magnetic field strength felt by the particle is $B_{\rm x}(y_{\rm max})=(y_{\rm max}/\delta)B_{0}$. Assuming that the electron remains in the radiation reaction limit within each cycle, the maximum energy reached increases as the magnetic field decreases such that $\gamma_{\rm e}=(\delta/y_{\rm max})\gamma_{\rm rad}$. The joint evolution of $\theta_0$, $y_{\rm max}$ and $\gamma_{\rm e}$ in this regime can be roughly estimated if one assumes that this situation is similar to the first regime described above where $B_0$ is replaced by $(y_{\rm max}/\delta)B_0$. Then, one finds $\theta_0\propto \gamma_{\rm e}^{-3/2}$.
\newline
\newline
It is important to note that a particle will not always undergo the three regimes in a sequence. This will depend on the initial conditions imposed on the electron and on the properties of the reconnection layer.

\begin{figure}
\epsscale{1.0}
\plotone{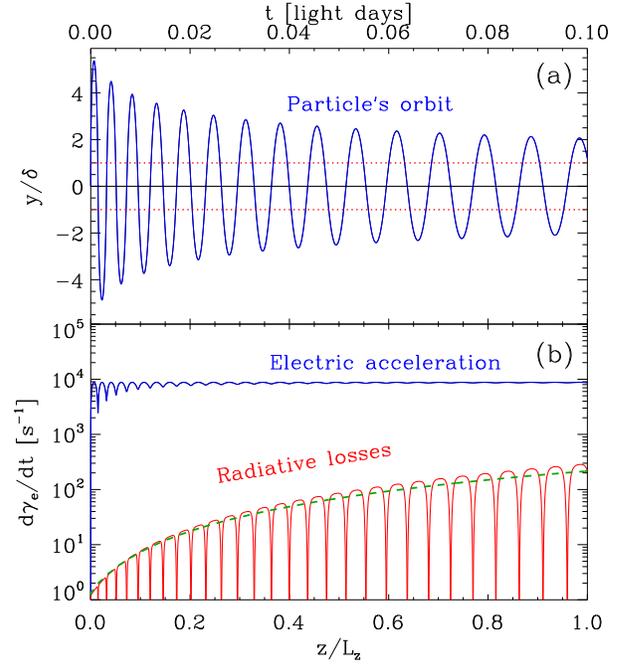}
\caption{Example of a numerically calculated relativistic Speiser orbit (blue solid line in panel (a) of an electron initially injected in the $yz$-plane at the origin of the axes with a Lorentz factor $\gamma_{\rm inj}=5\times 10^6$ and $\theta_{\rm inj}=90\degr$. Outside the layer, the reconnecting magnetic field is $B_0=5~$mG. The electric field is uniform with $\beta_{\rm rec}=0.1$ and the length of the layer $L_{\rm z}=0.1$~light-day. The red dotted lines delimit the thickness of the reconnection layer, $y=\pm \delta=\pm 10^6\rho_0$. The bottom panel (b) compares the evolution of the radiative losses (red line) with the almost constant electric acceleration rate (blue line). The green dashed line is the radiative loss averaged over each cycle. The reconnected $B_{\rm y}$ and the guide $B_{\rm z}$ fields are neglected.}
\label{fig_orb}
\end{figure}

\begin{figure}
\epsscale{1.0}
\plotone{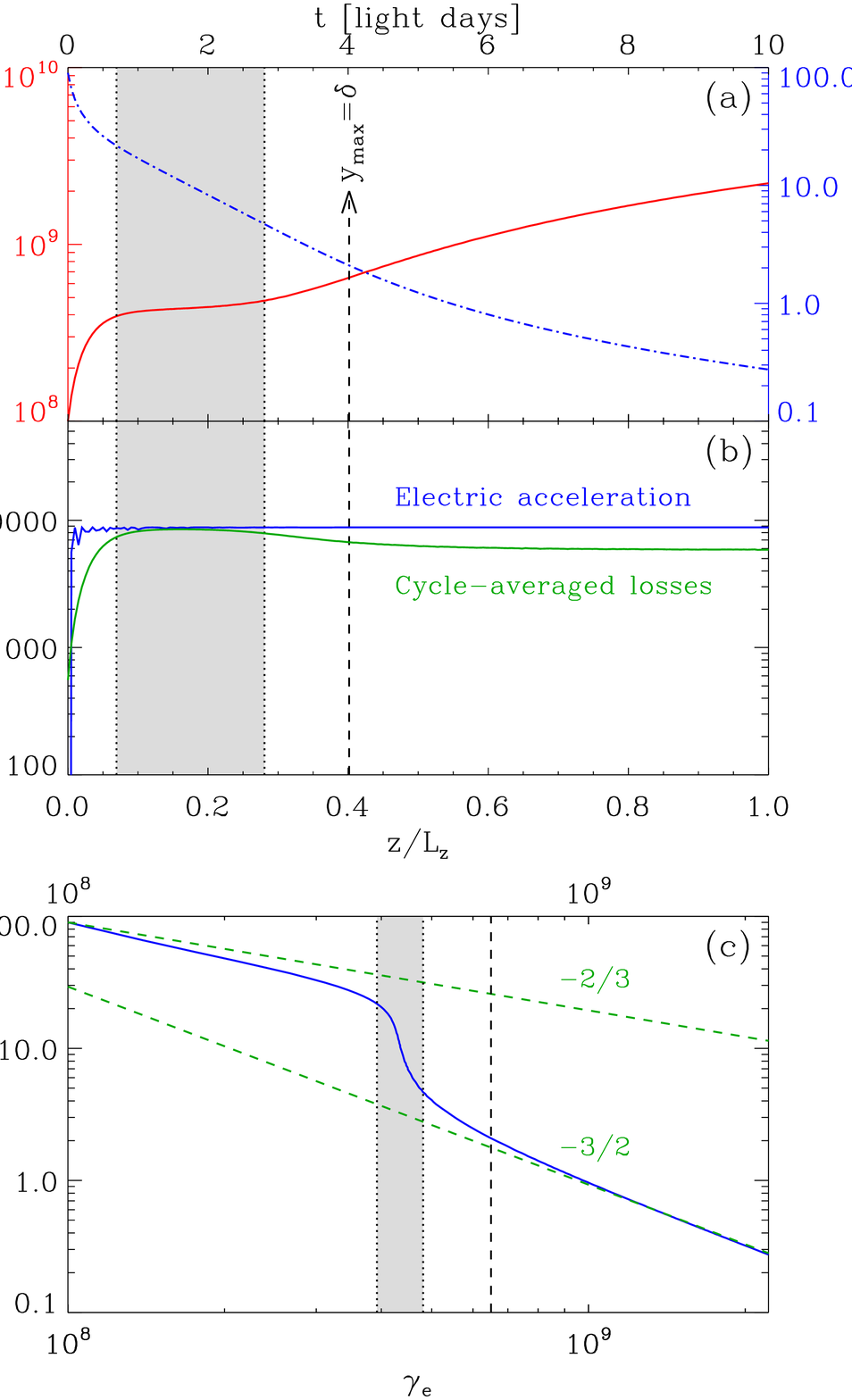}
\caption{\emph{Panel (a)}: Evolution of the electron's Lorentz factor $\gamma_{\rm e}$ (red solid line) and the midplane-crossing angle $\theta_0$ (blue dotted-dashed line) along the orbit inside the reconnection layer of length $L_{\rm z}=10$~light days $=3\times 10^{16}~$cm and thickness $\delta=10^6\rho_0=3.4\times 10^{11}$~cm. \emph{Panel (b)}: Evolution of the electric acceleration rate (blue line) and the cycle-averaged radiative losses (green line). \emph{Panel (c)}: Relation between $\theta_0$ and $\gamma_e$ calculated along the orbit. The green dashed lines give the analytical solutions derived in Section~\ref{analytical} (power-laws of index $-2/3$ and $-3/2$). In all panels, the dashed vertical line indicates the place where the orbit of the electron lies completely inside the current layer ($y_{\rm max}\lesssim\delta$). The gray region show the period where the electron is in the radiation reaction limit regime ($\gamma_{\rm e}=\gamma_{\rm rad}$). $B_0=5~$mG, $\beta_{\rm rec}=0.1$ and the reconnected and the guide fields are neglected as well as Compton drag ($\mathcal{U}_{\star}$=0).}
\label{fig_evol}
\end{figure}

\subsection{Comparison with numerical solutions}\label{compare}

\begin{figure*}
\epsscale{1.0}
\plotone{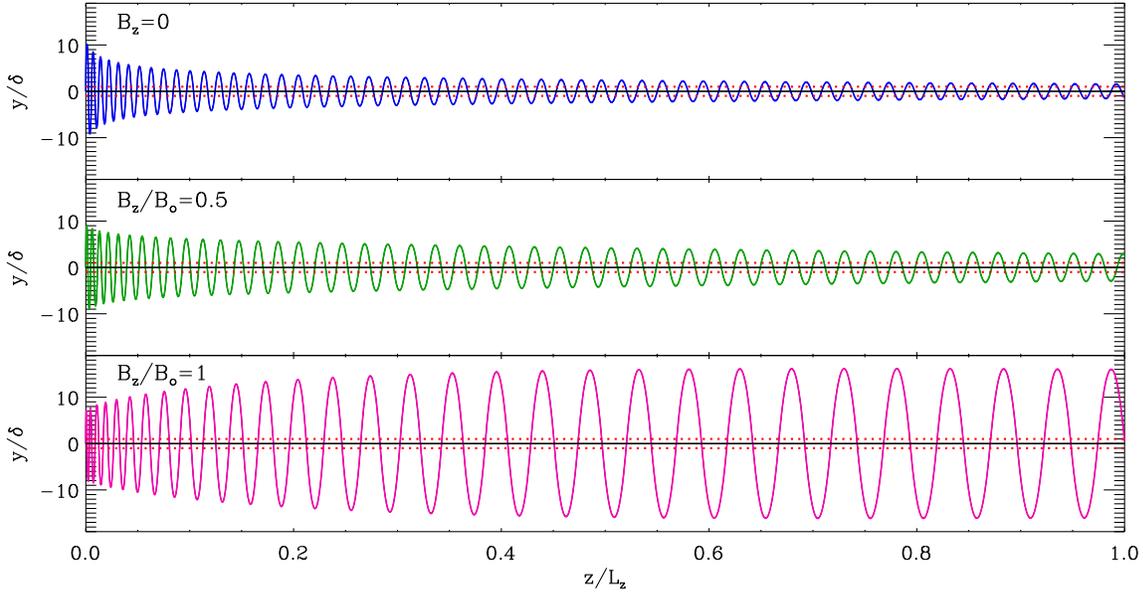}
\caption{Examples of electron orbits projected in the $yz$-plane for a finite guide field $B_{\rm z}$=0 (\emph{top}), 0.5~$B_0$ (\emph{middle}),~and $B_0$ ({\em bottom}). The particle is initially injected with $\gamma_0=10^7$ at the origin along the $y$-axis ($\theta_0=90\degr$). The layer is $L_{\rm z}=3\times 10^{15}~$cm (1-light day) long, of thickness $\delta=3.4\times 10^{11}$~cm, with $B_0=5~$mG, $\beta_{\rm rec}=0.1$, $B_{\rm y}=0$ and $\mathcal{U}_{\star}$=0.}
\label{fig_bz}
\end{figure*}

\begin{figure}
\epsscale{1.0}
\plotone{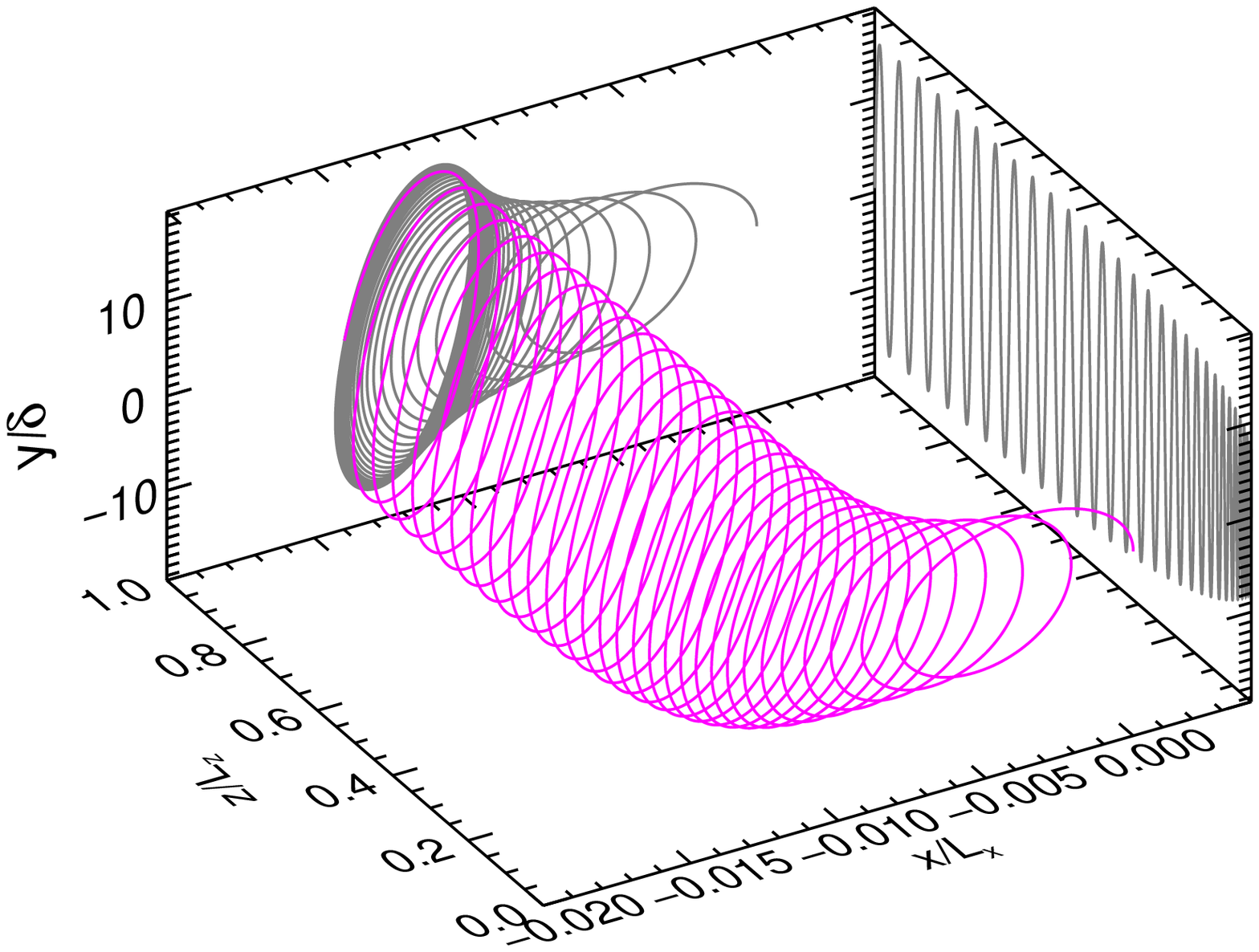}
\caption{Three-dimensional electron orbit corresponding to the case where $B_{\rm z}=B_0$ in Fig.~\ref{fig_bz}. The grey lines show the projections of the orbit in the $yz$- and $xy$-planes.}
\label{fig_3d}
\end{figure}

In order to track the particle's orbit accurately over many cycles across the reconnection layer, a high-order numerical scheme is required. We use the explicit Runge-Kutta-Verner procedure of the eighth order \citep{1978SJNA...15..772V}. Figure~\ref{fig_orb} shows an example of a numerically integrated orbit of an electron evolving in a uniform electric field $E_{\rm z}=0.1~B_0$ with $B_0=5~$mG and the reconnecting magnetic field $B_{\rm x}$ reversing over a layer of finite thickness $\delta=3.4\times 10^{11}$~cm (corresponding to $\delta=\rho_{\rm e}(\gamma_{\rm e}=10^6)$). The reconnected $B_{\rm y}$ and the guide $B_{\rm z}$ fields are neglected at this stage in order to compare with the above analytical solutions. As expected, we find that the orbit shrinks steadily toward the midplane.

Figure~\ref{fig_evol} shows the long-term evolution of the particle's orbit parameters $\gamma_e$ and $\theta_0$. The initial parameters of the electron were chosen specifically to obtain all three regimes described in the previous section in a sequence. At first (regime 1, $z\lesssim 0.75$~light-day, 1 light-day $\approx 3\times 10^{15}~$cm), most of the electron's orbit is well outside the layer ($y_{\rm max}\gg \delta$, $\delta=3.4\times 10^{11}$~cm) and the electron undergoes almost pure electric acceleration ($\gamma_{\rm e}\ll\gamma_{\rm rad}$). The midplane crossing angle almost follows the power-law $\theta_0\propto\gamma_e^{-2/3}$. The deviations from the analytic solution are entirely attributed to the effect of the radiation reaction force, not completely negligible here. Then, the electron enters the radiation-reaction limit regime (regime 2, $0.75\lesssim z/(1~{\rm light~day})\lesssim 2.75$, gray shaded area) where the Lorentz factor saturates at $\gamma_{\rm rad}\approx 4.3\times 10^8$ and $\theta_0$ decreases exponentially (linearly on a linear-logarithmic scale), in agreement with the analytical model. After some time, the orbit shrinks inside the layer (regime 3, $y_{\rm max}<\delta$, $z>1.2\times 10^{16}~$cm). Radiative losses no longer counterbalance the electric acceleration exactly over each cycle and the ratio of these two quantities tends to about $\sim 2/3$. The electron is continuously focused and accelerated in the $z$-direction to energies above the standard radiation reaction limit associated with the $B_0$ field ($\gamma_{\rm e}>\gamma_{\rm rad}$), and $\theta_0$ decreases asymptotically with energy as $\theta_0\propto \gamma_{\rm e}^{-3/2}$.

\subsection{The guide field}\label{sect_bz}

In addition to $B_{\rm x}$ and $E_{\rm z}$, we study in this section the effect of a uniform finite guide field $B_{\rm z}$ and the induced in-plane electric fields $E_{\rm x}$ and $E_{\rm y}$ on the motion of the particle. The reconnected field $B_{\rm y}$ and magnetic islands are still neglected at this stage. Figure~\ref{fig_bz} presents a few numerically integrated orbits projected onto the $yz$-plane where only the guide field strength changes from $B_{\rm z}=0$ to $B_{\rm z}=0.5~B_0$ and $B_{\rm z}=B_0$. We find that the presence of a strong guide field ($B_{\rm z}\gtrsim B_0$) suppresses the focusing of the particle toward the reconnection layer, even if the particle is injected inside the layer at a small angle, $\theta_0\ll 1$. For $B_{\rm z}=B_0$, the particle's orbit is still tied to the layer (the particle does not escape in the $y$-direction) but the particle remains well outside of the layer ($y_{\rm max}\gg\delta$) most of the time where its energy is limited by the radiation reaction force. For a moderate guide field strength ($B_{\rm z}=0.5~B_0$), the particle focuses toward the midplane as in the case without guide field, but at a slower rate (see the middle panel in Figure~\ref{fig_bz}).

Examination of the three-dimensional orbit for $B_{\rm z}=B_0$ shows that the particle spirals along the guide field in the $xy$-plane and drifts slightly in the $x$-direction (Fig.~\ref{fig_3d}). This drift in the $x$-direction as well as the defocusing of the particle's orbit in the $yz$-plane in a strong guide field is mainly due to the in-plane electric field $E_{\rm x}$. Indeed, $E_{\rm x}$ accelerates the particle coherently in the $x$-direction as the particle makes one turn along the guide field in the $xy$-plane, resulting in an increase of the Larmor radius of the particle. In contrast, $E_{\rm y}$ results only in a small net energy gain because it first decelerates and then accelerates the particle in the $y$-direction by almost the same amount during one cycle along $B_{\rm z}$ (depending on where the particle is in the $xy$-plane, see Fig.~\ref{fig_electric}). To summarize this section, we find that a strong guide field $B_{\rm z}\gtrsim B_0$, does not allow the particles to converge into the layer. As a result, particles cannot be accelerated above the standard radiation reaction limited energy $\gamma_{\rm rad}$.

\subsection{The reconnected field and magnetic islands}\label{sect_by}

\begin{figure}
\epsscale{1.0}
\plotone{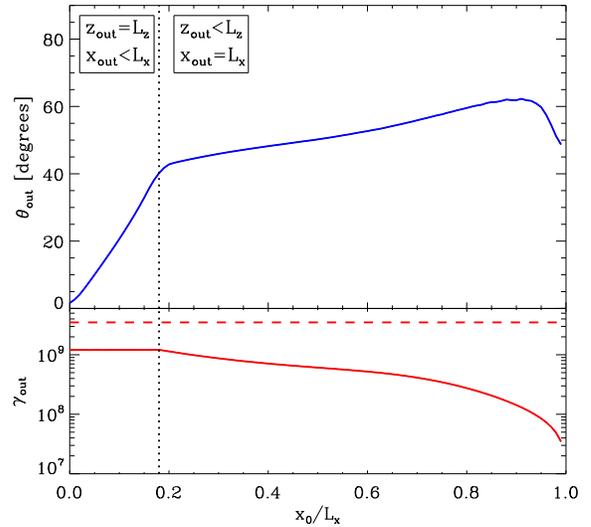}
\caption{Lorentz factor ({\em bottom} panel) and deviation angle (defined with respect to the $z$-axis, {\em top} panel) of the electron at the end of the reconnection layer as a function of the initial position of the particle inside the layer $x_0\leq L_{\rm x}$. Initially, the electron is injected along the $y$-axis ($\theta_0=90\degr$) with a Lorentz factor $\gamma_0=10^7$, inside a layer of length $L_{\rm z}=2L_{\rm x}=1.2\times 10^{16}~$cm (4-light days) and thickness $\delta=3.4\times 10^{11}~$cm, with $B_0=5~$mG and $\beta_{\rm rec}=0.1$. The vertical dotted line separates the region where the particle escapes the layer in the $z$-direction first ($x_0/L_{\rm x}<0.18$) from the region where it escapes in the $x$-direction first ($x_0/L_{\rm x}>0.18$). The horizontal red dashed line marks $\gamma_{\rm max}=e\beta_{\rm rec}B_0 L_{\rm z}/(m_{\rm e}c^2)$. There are no magnetic islands and $B_{\rm z}=0$ in this calculation.}
\label{fig_by}
\end{figure}

We next investigate the effect of the reconnected magnetic field $B_{\rm y}$ on the motion of the particles in the vicinity of the layer. We will first assume that $B_{\rm y}$ follows the linear law given by Eq.~(\ref{by}) and magnetic islands are still neglected at this stage. To single out the effect of $B_{\rm y}$ only, there is no guide field and no Compton drag in this subsection. All fields are fully included later in the population studies (Section~\ref{pop}).

In Figure~\ref{fig_by}, we calculate the Lorentz factor of the particle $\gamma_{\rm out}$ and the angle of deviation of the orbit $\theta_{\rm out}$ from the $z$-axis when the particle reaches the boundaries of the layer ({\em i.e.} when $z=L_{\rm z}$ or $|x|=L_{\rm x}$), as a function of the initial injection distance $x_0$ (between $x_0=0$ and $x_0=L_{\rm x}$, $y_0=0$ and $z_0=0$). The initial particle velocity is along the $y$-axis with a Lorentz factor $\gamma_0=10^7$. The layer is $L_{\rm z}=2L_{\rm x}=4$~light-day-long and $\delta=3.4\times 10^{11}$~cm thick, with $B_0=5$~mG. As expected, we find that the reconnected magnetic field $B_{\rm y}$ deflects the particle in the $x$-direction away from the $z$-axis. The deviation angle $\theta_{\rm out}$ increases quasi-linearly with the injection distance $x_0$ because of the linear increase of $B_{\rm y}$ with $x$. The break at $x_{0, \rm bk}/L_{\rm x}\approx 0.18$ is geometrical in origin because it distinguishes whether the particle escapes the layer at $z=z_{\rm out}=L_{\rm z}$ or at $x=x_{\rm out}=L_{\rm x}$. Below $x_{0, \rm bk}$, the particle exits the layer in the $z$-direction first ($x_{\rm out}<L_{\rm x}$), and above $x_{\rm bk}$ the particle is deflected out of the layer in the $x$-direction first ($z_{\rm out}<L_{\rm z}$). This explains why the particle is accelerated to the highest energy if $x_0<x_{0,\rm bk}$ because it goes through the total electric potential drop available, $e\beta_{\rm rec}B_0L_{\rm z}\approx 2~$PeV (see Fig.~\ref{fig_by}, \emph{bottom} panel). In addition, we observe that the presence of the reconnected field $B_{\rm y}$ does not affect the focusing mechanism of the particle's orbit toward the midplane, but instead of having a beam of particles focused in the $z$-direction, the trajectories spread over a wide range of angles $\left|\theta_{\rm out}\right|\lesssim 90\degr$ in the $xz$-plane at the end of the reconnection layer (fan beam of particles).

\begin{figure}
\epsscale{1.0}
\plotone{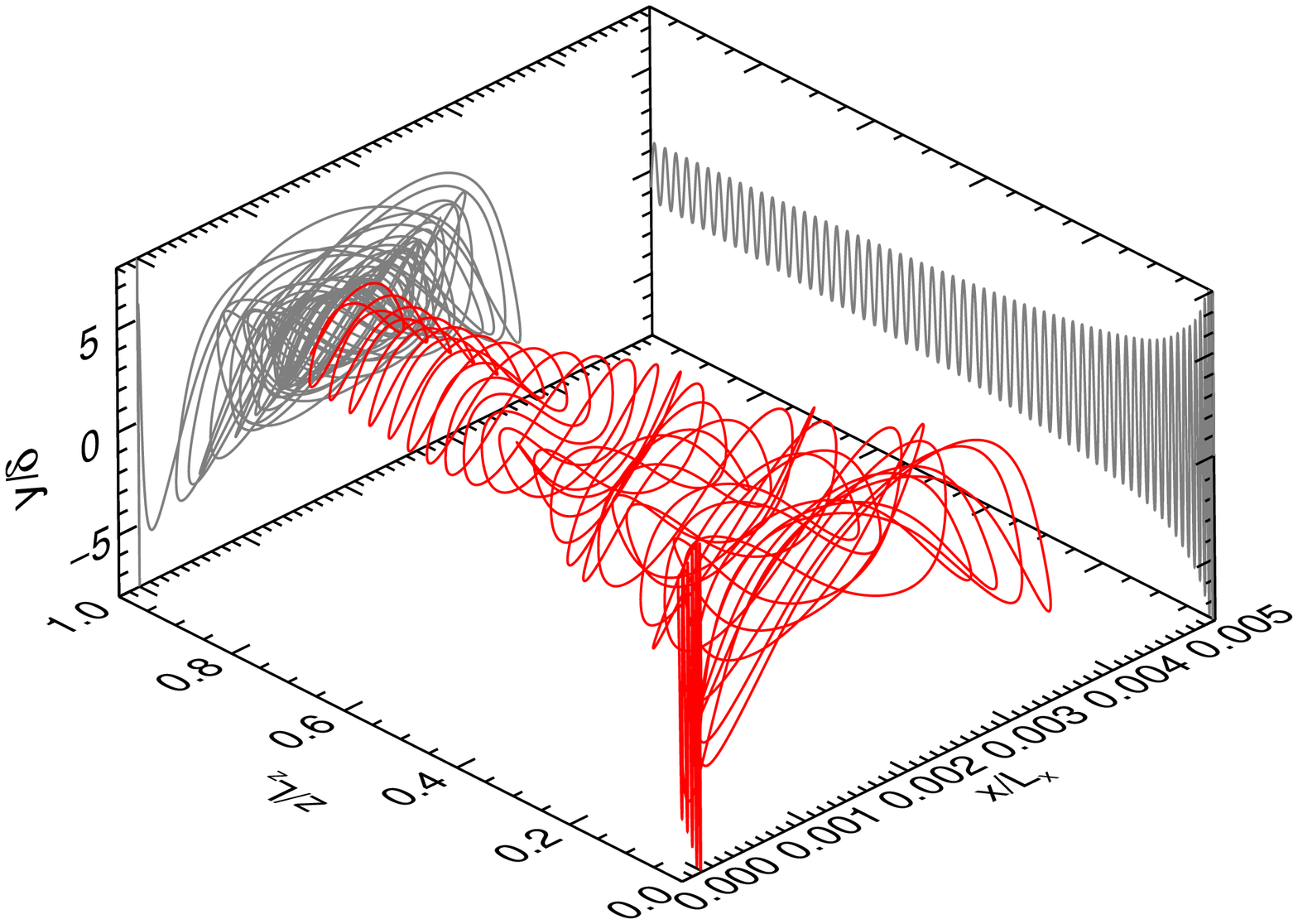}
\plotone{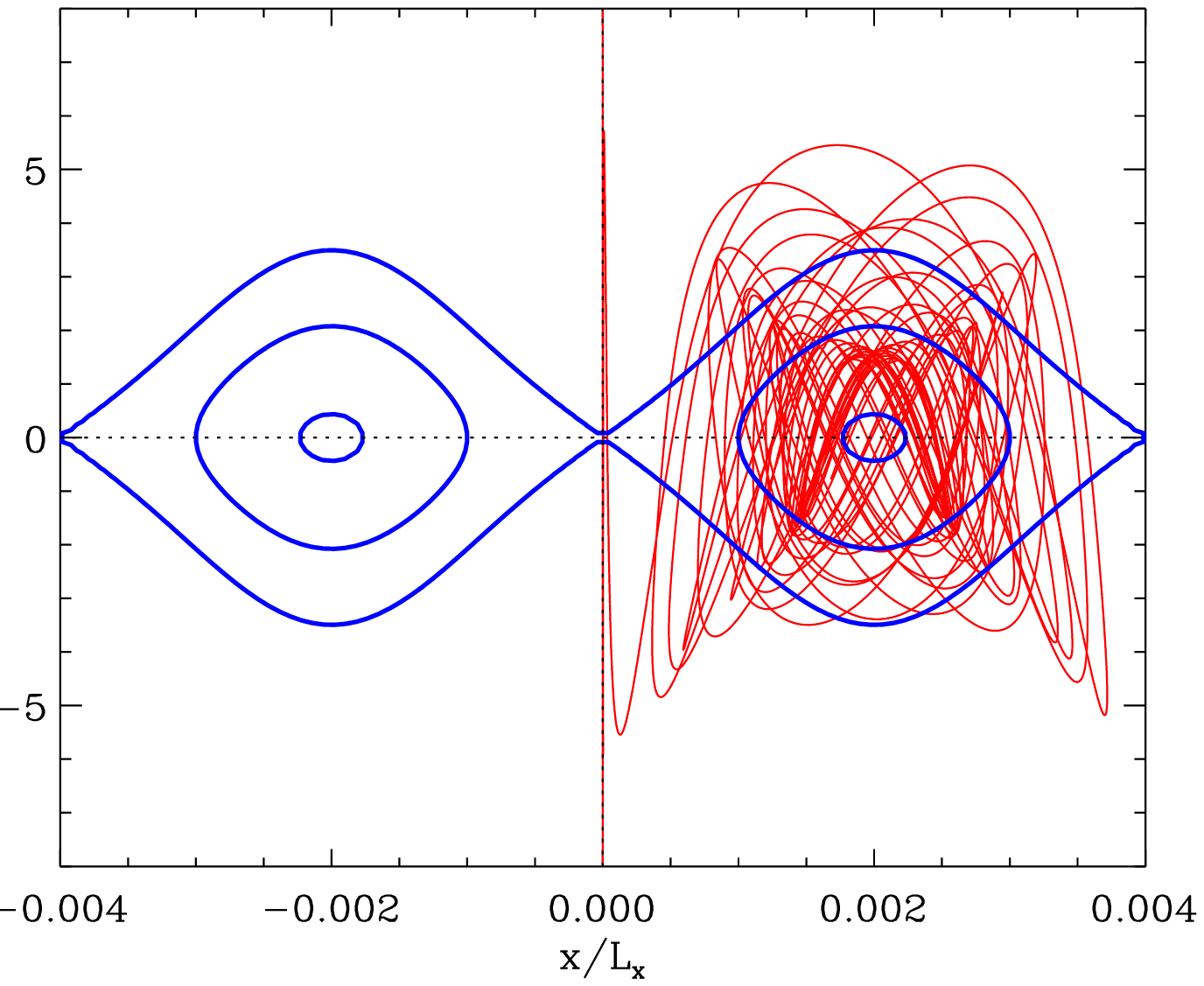}
\caption{{\em Top panel:} Example of a three-dimensional orbit of a particle trapped in a magnetic island present in the reconnection layer. The layer is $L_{\rm z}=1~$light-day-long, $\delta=3.4\times 10^{11}~$cm with $B_0=5~$mG and contains a static chain of $n_{\rm is}=500$ islands with $\epsilon=1.4$. The particle is injected at the origin along the $y$-axis with $\gamma_0=10^7$. There is no guide field. The grey lines show the projections of the orbit in the $yz$- and $xy$-planes.{\em Bottom panel:} Projection of the orbit in the $xy$-plane overlaid on the magnetic field lines of two islands in the layer.} 
\label{fig_is}
\end{figure}

As discussed in Section~\ref{reconnection}, a thin current layer is unstable to tearing modes, leading to the formation of a dynamical chain of secondary magnetic islands (where fields lines are closed loops: see Fig.~\ref{fig_fields}, {\em bottom} panel, for an illustration). In order to quantify the impact of magnetic islands in the layer on the dynamics of the particles, we add to the linearly growing reconnected field $B_{\rm y}$ a sinusoidal perturbation $\tilde{B}_{\rm y}$ prescribed in Eq.~(\ref{by_is}) (see the discussion in Section~\ref{reconnection}). Islands are static and identical flux ropes, of length $L_{\rm z}$ in the $z$-direction and of width $W_{\rm x}=2L_{\rm x}/n_{\rm is}$ in the $x$-direction. In the $y$-direction, the width of each island is about $W_{\rm y}\sim\epsilon\delta$ if $\epsilon\gtrsim 1$, and $W_{\rm y}\sim (2\epsilon)^{1/2}\delta$ if $\epsilon\lesssim 1$.

Calculations show that in some cases, the particle is trapped within one island and its orbit stays aligned along the $z$-direction. In the $xy$-plane, the particle's orbit is confined and focuses toward the center of the island (O-point). Figure~\ref{fig_is} shows that the three-dimensional trajectory of the particle becomes quite complex and it is rapidly trapped by one island. For this calculation, we considered the same particle ($\gamma_0=10^7$) as in Figure~\ref{fig_by} with $x_0=0$ for a $L_{\rm z}=1~$light-day-long layer, $\delta=3.4\times 10^{11}~$cm and $B_0=5~$mG. The layer contains $n_{\rm is}=500$ islands and $\epsilon=1.4$ so that $\tilde{B}_{\rm y,max}=0.5~B_0$. This magnetic confinement by the islands operates if the large scale reconnected field (linearly increasing with $x$) is locally smaller than the amplitude of the island field $B_{\rm y}\lesssim\tilde{B}_{\rm y,max}=\epsilon n_{\rm is}\pi B_0 (\delta/L_{\rm x})$, {\em i.e.} if the particle is injected at a distance $x_0\lesssim \epsilon\pi n_{\rm is} \delta/\beta_{\rm rec}$. Otherwise, the particle feels only the large scale magnetic deflections by $B_{\rm y}$. If the amplitude $\tilde{B}_{\rm y, max}$ exceeds the maximum large scale reconnected field $B_{\rm y,max}=\beta_{\rm rec}B_0$, then we find that the particle is systematically confined within an island for any value of $x_0$, and regardless of the size of the island with respect to the Larmor radius of the particle. We conclude that the presence of magnetic islands in the current sheet does not hamper the acceleration of the high-energy particles of interest here; they even help in keeping particles aligned along the electric field. Note however that this simple model is static, {\em i.e.} it does not take into account any electric fields associated with the contraction or the merging of islands, which could affect particle acceleration in a similar way as for smaller particle energies \citep{2006Natur.443..553D}.

\begin{figure*}
\epsscale{1.0}
\plotone{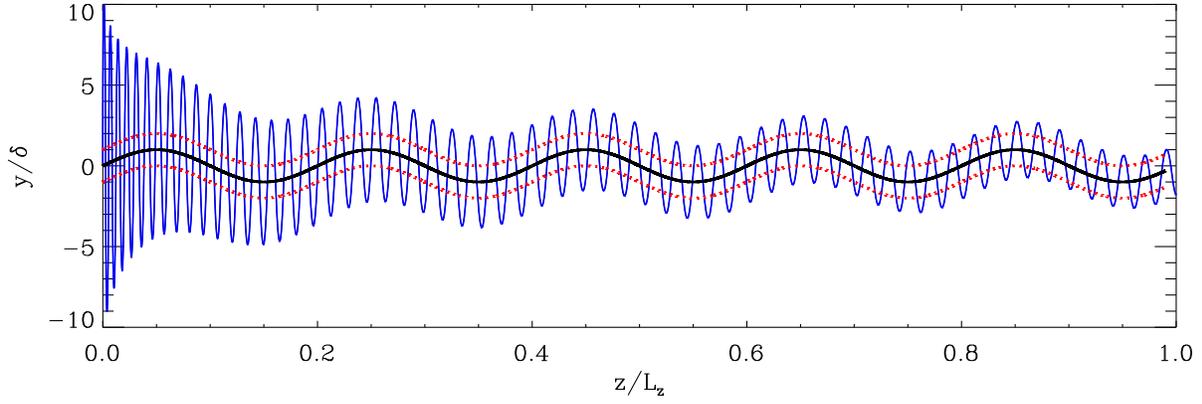}
\caption{An electron trajectory in the $yz$-plane in a current layer subject to a sinusoidal deformation to mimic the effect of a kink-like instability in the layer, with $C_{\rm y}=1$ and $n_{\rm z}=5$. The location where $B_{\rm x}=0$ is shown by the black solid line and the distance $\pm\delta$ to this line is indicated by the red dashed lines. Apart from $B_{\rm x}$, all the other parameters of the layer and of the particle are the same as in Fig.~\ref{fig_bz} but with no guide field.}
\label{fig_kink}
\end{figure*}

\subsection{Kink instability}\label{sect_kink}

We now consider a sinusoidal deformation of the current layer in the $y$-direction as a function of $z$, using the expression in Eq.~(\ref{eq_kink}) for the reconnecting field to mimic the effect of a kink instability (during the linear or early nonlinear development). The calculation of particle trajectories shows that the deformation of the layer, even with high amplitudes ($C_{\rm y}\gtrsim 1$), does not change at all the focusing mechanism. The final energy reached by the particle at the end of the layer is also unchanged. Figure~\ref{fig_kink} is an example of a particle trajectory initially injected at the origin along the $y$-direction with $\gamma_0=10^7$ in a layer with $C_{\rm y}=1$ and $n_{\rm z}=5$. In this calculation, the layer is $L_{\rm z}=1~$light-day long and $\delta=3.4\times 10^{11}~$cm thick with $B_0=5~$mG and $\beta_{\rm rec}=0.1$. There are no magnetic islands, guide field or external photon fields for simplicity. The layer guides the particle trajectory. The kink affects the energy of the particle and the focusing mechanism only if the wavelength of the kink mode considered $\lambda_{\rm z}=L_{\rm z}/n_{\rm z}$ becomes shorter than the distance traveled by the particle along the $z$-direction during one half Speiser cycle $z_1$ (see Sect.~\ref{analytical}). In this case ($\lambda_{\rm z}\ll z_1$), the particle always feels a non-zero average perpendicular magnetic field during each cycle. Its energy becomes limited by the radiation reaction force and the particle never reaches the inside of the layer.

\subsection{Inverse Compton cooling}\label{effect-compton}

\begin{figure}
\epsscale{1.0}
\plotone{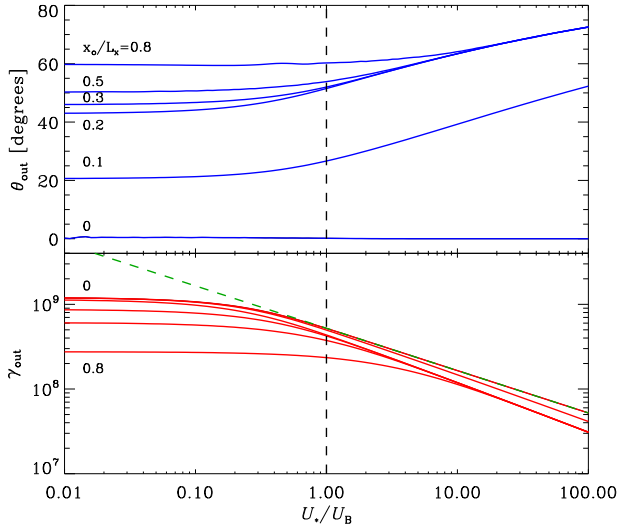}
\caption{Final Lorentz factor ({\em bottom} panel) and deviation angle ({\em top} panel) of the particle's trajectory as functions of the ratio between the photon energy density and the magnetic energy density $\mathcal{U}_{\star}/\mathcal{U}_{\rm B}$. Each line corresponds to different initial distance $x_{0}/L_{\rm x}$=0, 0.1, 0.2, 0.3, 0.5, and 0.9. The initial conditions and the properties of the reconnection layer are the same as in Fig.~\ref{fig_by}. The green dashed line is the solution given in Eq.~(\ref{grad_ic}).}
\label{fig_ic}
\end{figure}

The inverse Compton drag force strongly affects the trajectory of the electron if the radiation energy density is of order the magnetic energy density $\mathcal{U}_{\star}\gtrsim \mathcal{U}_{\rm B}\equiv B_0^2/8\pi$. To quantify the effect of inverse Compton drag, we compute the deviation angle $\theta_{\rm out}$ and the Lorentz factor $\gamma_{\rm out}$ of the particle at the end of the layer as a function of the ratio $\mathcal{U}_{\star}/\mathcal{U}_{\rm B}$. Figure~\ref{fig_ic} shows the result of this calculation where the particle is injected at different locations $x_0$ in the midplane ($y_0=0$ and $z_0=0$). The layer has the same properties as in Section~\ref{sect_by}, with $B_{\rm z}=0$ and no magnetic islands. For simplicity, the background of target photons is homogeneous and isotropic in the reconnection region. We find that inverse Compton cooling increases the deflection of the particle's trajectory by the reconnected field $B_{\rm y}$ in the $x$-direction (except for electrons injected very close to the $z$-axis). This is due to the efficient counterbalance of the electric acceleration rate by the Compton energy losses. High Compton cooling has a more critical effect on the energy of the electrons at the end of the layer. If synchrotron losses are negligible ($\mathcal{U}_{\star}\gg B_0^2/8\pi$) and if deviations in the $x$-direction are not too important ($x_0/L_{\rm x}\ll 1$), the final energy of the electron is given by the balance between the electric force and the Compton drag force, {\em i.e.} (in the Thomson regime)
\begin{equation}
\gamma_{\rm rad}^{\rm ic}=\left(\frac{3 e \beta_{\rm rec} B_0}{4\sigma_{\rm T}\mathcal{U}_{\star}}\right)^{1/2}.
\label{grad_ic}
\end{equation}
This expression is the equivalent of the standard radiation reaction limited energy derived in Eq.~(\ref{grad}) in the case where the Compton drag force dominates over the radiation reaction force. High inverse Compton cooling severely limits the acceleration of the particle to extreme energies, but we note that this does not affect the concentration of the particle toward the reconnection layer. In fact, the Compton drag force helps in focusing the orbit toward the $xz$-plane.

\section{Population studies}\label{pop}

\begin{figure}
\epsscale{1.0}
\plotone{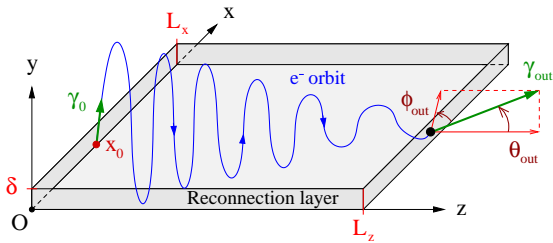}
\caption{This diagram depicts qualitatively the motion of a single electron in the layer, injected at some distance $x_0$ from the origin of the axis with a Lorentz factor $\gamma_0$. At the end of the layer ($z=L_{\rm z}$ or $|x|=L_{\rm x}$), the Lorentz factor of the electron is $\gamma_{\rm out}$ and its direction of motion is given by the spherical angles $\theta_{\rm out}$ (angle to the $z$-axis) and $\phi_{\rm out}$ (angle to the $x$-axis in the $xy$-plane).}
\label{geo_pop}
\end{figure}

\begin{figure*}
\epsscale{1.0}
\plotone{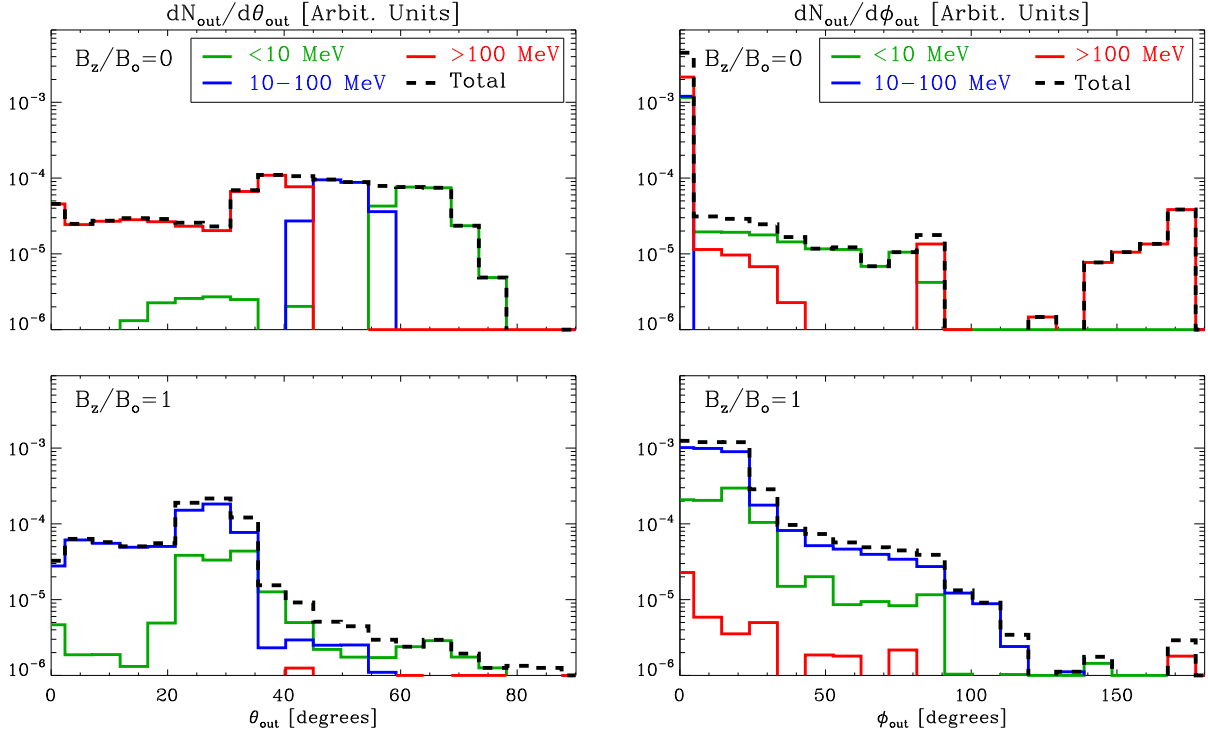}
\caption{Outgoing angular distributions ${\rm d}N_{\rm out}/{\rm d}\theta_{\rm out}$ ({\em left} panels) and ${\rm d}N_{\rm out}/{\rm d}\phi_{\rm out}$ ({\em right} panels) of a population of particles ($N_{\rm part}=97,200$) initially injected with a power-law of index $-2$ with $10^6<\gamma_0<10^9$ and with an isotropic angular distribution. All particles start at the begining of the layer ($y(0)=0$, $z(0)=0$) and are uniformly spread along the $x$-axis between $0<x_0<L_{\rm x}$, where $2 L_{\rm x}=L_{\rm z}=4$~light-days. The guide field is $B_{\rm z}=0$ ({\em top} panel), and $B_{\rm z}=B_0$ ({\em bottom} panel), where $B_0=5~$mG. The reconnection rate is $\beta_{\rm rec}=0.1$, and Compton losses are neglected. Each line indicates the distribution of particles emitting synchrotron photons in the following energy bands: $<10~$MeV (green), 10-100 MeV (blue), $>100~$MeV (red) and the sum of all bands (dashed line), assuming that electrons are bathed in the magnetic field $B=B_0$.}
\label{fig_ang}
\end{figure*}

\begin{figure}
\epsscale{1.0}
\plotone{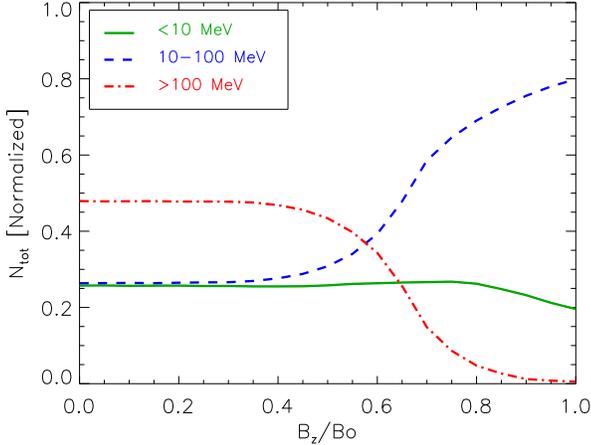}
\caption{Effect of the guide field strength on the outgoing distribution of particles emitting synchrotron radiation in the energy bands $<10~$MeV (green solid line), $10-100~$MeV (blue dashed line) and >100 MeV (red dotted-dashed line), for electrons bathed in the magnetic field $B=B_0$.}
\label{fig_nbz}
\end{figure}

\begin{figure}
\epsscale{1.0}
\plotone{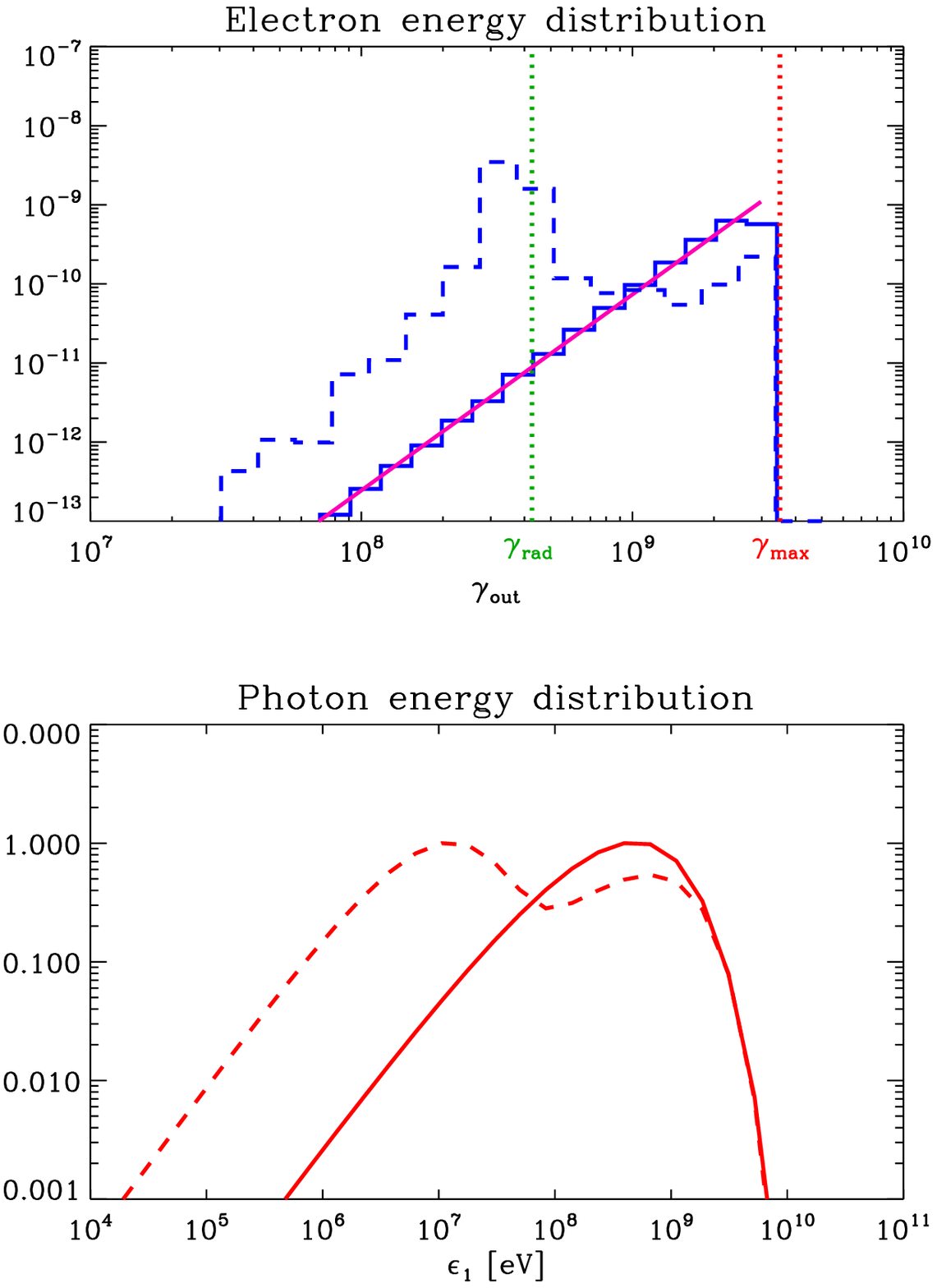}
\caption{Outgoing energy distribution of the same population of electrons ({\em top} panel) accelerated in the same reconnection layer as described in Fig.~\ref{fig_ang} for $B_{\rm z}=0$ (blue solid line), and $B_{\rm z}=B_0$ (blue dashed line). A power-law fit to the electron spectrum (with no guide field) ${\rm d}N/{\rm d}\gamma_{\rm e}\propto \gamma_{\rm e}^{0.5}$ is overlaid (magenta solid line). The maximum energy obtained by pure linear acceleration $\gamma_{\rm max}=eE_0L_{\rm z}/m_{\rm e}c^2=3.5\times 10^9$ is shown by the red vertical dotted line. The green vertical dotted line indicates the radiation reaction limited energy $\gamma_{\rm rad}$ in $B_0$. The resulting synchrotron emission spectra at the end of the layer are also shown ({\em bottom} panel) assuming that the particles evolve in a disordered magnetic field of strength $B=B_0$.}
\label{fig_spec}
\end{figure}

This section is dedicated to the study of dynamics of a population of particles injected in the reconnection layer.

\subsection{Simulation setup}

We perform an exhaustive exploration of the parameter space composed of the initial conditions of the electron, namely the Lorentz factor $\gamma_0$, the direction of motion defined by the spherical angles $\theta_0=[0,\pi]$ (angle with respect to the $z$-axis) and $\phi_0=[0,2\pi]$ (angle with respect to the $x$-axis in the $xy$-plane), and the coordinates $x_0=[0,L_{\rm x}]$ and $y_0=[0,\delta]$ for $z_0=0$. For each set of initial parameters, the orbit of the electron is calculated following the numerical procedure described in Section~\ref{compare}, and the value of each of these parameters at the end of the layer is stored in a table and is denoted with a subscript ``out''. Each particle is followed until it crosses the outer boundary of the layer, {\em i.e.} $|x|=L_{\rm x}$ or $z=L_{\rm z}$ (see Fig.~\ref{geo_pop}). In practice, $y_0$ is set to 0 because the final result does not depend much on the value of this parameter (if $\left|y_0\right|<\delta$).

From this calculation, we can derive the distributions of the accelerated electrons at the end of the reconnection layer, in particular the angular and energy distributions. Initially, the electrons have isotropic orientations and are uniformly distributed in space (along $x$) and follow a power-law energy distribution of index $p=2$ such that $dN_0/d\gamma_0\propto\gamma_0^{-p}$ with $\gamma_1<\gamma_{0}<\gamma_2$. The low-energy cut-off is fixed at $\gamma_1=10^6$ and corresponds to the Lorentz factor of the bulk particles governing the thickness of the layer $\delta=3.4\times 10^{11}~$cm in $B_0=5~$mG field strength. The high-energy cut-off is set at $\gamma_2=10^9$, {\em i.e.} for the particles with a Larmor radius of order the width of the layer $L_{\rm x}$ in the $B_{\rm y}(|x|=L_{\rm x})=\beta_{\rm rec}B_0$ magnetic field. In these calculations, all large-scale fields are present with a uniform $E_{\rm z}=\beta_{\rm rec}B_0$, $B_{\rm x}$ following Eq.~(\ref{bx}), $B_{\rm y}$ following Eq.~(\ref{by}) but with no magnetic islands or curvature in the layer, and a uniform guide field $B_{\rm z}=0,~$or $1$ times $B_0$, with the associated in-plane electric field components $E_{\rm x}$ and $E_{\rm y}$. We follow the trajectories of 97,200 particles for each simulation.

\subsection{Outgoing particle angular and energy distributions}

Figure~\ref{fig_ang} shows the outgoing angular distributions of particles with respect to $\theta_{\rm out}$, ${\rm d}N_{\rm out}/{\rm d}\theta_{\rm out}$ summed over $\phi_{\rm out}$, and with respect to $\phi_{\rm out}$, ${\rm d}N_{\rm out}/{\rm d}\phi_{\rm out}$ summed over $\theta_{\rm out}$. The distribution ${\rm d}N_{\rm out}/{\rm d}\theta_{\rm out}$ is roughly flat between $\theta_{\rm out}=0\degr$ and $90\degr$ (the distribution exhibits a slight excess between $\theta_{\rm out}=30\degr$ and $70\degr$) whereas ${\rm d}N_{\rm out}/{\rm d}\phi_{\rm out}$ is concentrated close to $\phi_{\rm out}=0\degr$, within a few degrees if $B_{\rm z}=0$. By symmetry with respect to the $y$-axis, if one injects a population of particles at $x_0=[-L_{\rm x},0]$, ${\rm d}N_{\rm out}/{\rm d}\theta_{\rm out}$ is distributed approximatively uniformly between $\theta_{\rm out}=-90\degr$ and $0\degr$, and ${\rm d}N_{\rm out}/{\rm d}\phi_{\rm out}$ is concentrated close to $\phi_{\rm out}=180\degr$. Hence, the reconnection layer transforms an initially isotropic distribution of particles into a fan beam in the $xz$-plane. This is associated with the focusing mechanism due to the alternating reconnecting magnetic field and electric field along the $z$-axis as discussed in detail in the previous sections. A more precise analysis indicates that about 90\% of the electrons, regardless of their energy, are contained within a solid angle $\Omega_{\rm out}\approx 0.1$ steradian, {\em i.e.} about 1\% of the full sphere. A flat distribution of the outgoing particles with respect to $\theta_{\rm out}$, ${\rm d}N_{\rm out}/{\rm d}\theta_{\rm out}\approx$ constant, implies that the number of particles per unit of solid angle ${\rm d}N_{\rm out}/\sin\theta_{\rm out}{\rm d}\theta_{\rm out}\propto\theta_{\rm out}^{-1}$ is highly concentrated along the $z$-axis, and is the relevant distribution to determine the observable emission.

In Figure~\ref{fig_ang}, we define three distinct populations of particles based on the characteristic synchrotron photon energy $\epsilon_{\rm syn}=(3/2)\gamma^2_{\rm out}\hbar e B_0/m_{\rm e} c$ that the particles would emit in the field $B_0$ once they leave the layer: $\epsilon_{\rm syn}<10~$MeV ($\gamma_{\rm out}<3.4\times 10^8$), $10~\rm{MeV}<\epsilon_{\rm syn}<100~{\rm MeV}$ ($3.4\times 10^8<\gamma_{\rm out}<10^9$), and $\epsilon_{\rm syn}>100~$MeV ($\gamma_{\rm out}>10^9$). We find that the majority of the electrons focused along the $z$-axis are also the most energetic ones. The low-energy particles ($\gamma_{\rm out}<10^9$) are dominant at high deflection angles $\theta_{\rm out}$ ($\theta_{\rm out}\gtrsim 45\degr$ for $B_{\rm z}=0$).

A strong guide field decreases the number of particles accelerated to the highest energies and broadens the angular spread of the particles at the end of the layer. Figure~\ref{fig_nbz} gives the total number of particles in each energy band as a function of the guide field strength. For small guide fields ($B_{\rm z}/B_0\lesssim 0.5$), the reconnection layer accelerates up to 50\% of the particles to the highest energies. Even for a moderately high guide field $B_{\rm z}/B_0=0.65$, the most energetic particles still represent 25\% of the total. However, if $B_{\rm z}$ is too strong, comparable to or greater than $B_0$, then only a few particles can be accelerated inside the reconnection layer to the highest energies (see the discussion in Section~\ref{sect_bz}). Most of the particles in this case spend most of their time outside the layer, and their energy is limited by the radiation reaction limit imposed by the external fields ({\em i.e.} $\epsilon_{\rm syn}<100~$MeV).

The upper panel in Figure~\ref{fig_spec} shows the overall energy distribution $\gamma^2_{\rm out} {\rm d}N_{\rm out}/{\rm d}\gamma_{\rm out}$ of all the electrons at the end of the layer for $B_{\rm z}/B_0=0,~$and 1. If there is no guide field, the spectrum is well fitted by a single power-law of index $+0.5$ (this result is robust against changes in the initial injection index $p=1$, $3$). The distribution extends almost up to the maximum energy that an electron can gain by pure electric linear acceleration $\gamma_{\rm max}=eE_0L_{\rm z}/m_{\rm e}c^2=3.5\times 10^9$. The distribution is quite narrow, almost monoenergetic. The highest energy particles at the end of the layer are mostly composed of the particles injected with a moderately high Lorentz factor $\gamma_0\sim 10^6$-$10^7$, because they have enough time to go deep inside the layer and they suffer little radiative losses initially. For high guide fields, the distribution differs from a single power-law because many particles pile-up at the radiation reaction-limited energy $\gamma_{\rm e}\approx 4\times 10^8$ (see Eq.~\ref{grad}), as mentioned previously.

\subsection{Resulting radiation spectrum}\label{pop_rad}

The lower panel in Figure~\ref{fig_spec} shows the resulting synchrotron spectrum emitted by all the particles. For a small guide field, it is close to the spectrum of a single electron with $\gamma_{\rm e}=\gamma_{\rm max}\approx3.5\times 10^9$, {\em i.e.} with frequency dependence of the flux $F_{\nu}\propto\nu^{1/3}$ and an exponential cut-off at the critical synchrotron frequency $\nu_{\rm c}=3 e B_{\perp}\gamma^2/4\pi m_{\rm e} c$ \citep{1970RvMP...42..237B}. For a strong guide field $B_{\rm z}=B_0$, the radiation spectrum split into two distinct synchrotron components. The low energy component peaks at around 10~MeV and corresponds to the synchrotron emission from the particles in the radiation reaction limit in $B_0=5~$mG field ($\gamma_{\rm rad}\approx 4\times 10^8$) and $\beta_{\rm rec}=0.1$. The high-energy component is fainter than the low energy one, peaks at about 1~GeV, and is emitted by the particles linearly accelerated to $\gamma_{\rm max}$ deep inside the layer.

The synchrotron spectrum is calculated assuming that the particles have escaped from the reconnection layer and entered a region of disordered magnetic field of strength $B_0=5~$mG (the spectrum is averaged over isotropic pitch angles). This spectrum represents the instantaneous emission right after the end of the layer, assuming that the layer ends abruptly\footnote{In reality, the layer may terminate gradually over a distance comparable to the length of the layer $L_{\rm z}$, and the particle would feel a perpendicular field strength smaller than $B_0$.}. For simplicity, this calculation does not take into account the photons emitted by the particles during their acceleration in the reconnection layer. However, this contribution should be small because most of the particles reach the maximum energy available in the electrostatic potential, indicating that the radiative losses are negligible during the acceleration. This statement is not true for a strong guide field ($B_{\rm z}\gtrsim B_0$), because the particles are stuck in the radiation reaction limit during most of their travel along the layer (Fig.~\ref{fig_spec}). In this case, the emission during the acceleration could be as important as or even greater than the radiation emitted at the end of the layer. The latter situation should not be considered in the context of the Crab Nebula flares, because the maximum energy of the photons emitted would not exceed $\sim160$~MeV contrary to what is observed. The spectrum emitted by a particle in the layer does not differ from synchrotron because the variation of the Lorentz factor of the particle is small over the formation length of a synchrotron photon, which is roughly given by the Larmor radius $\rho_{\rm e}$ divided by $\gamma_{\rm e}$ or $\rho_0=m_{\rm e}c^2/eB_0\sim 3.4\times 10^{5}~$cm. (Typically the particle gains a small amount of energy within one half Speiser cycle $z_1\gg\rho_0$, see Sect.~\ref{analytical}.)

Another important issue concerns the calculation of the radiation spectrum seen by an observer located in a given direction with respect to the reconnection layer. In the general case, one would need to know the precise structure of the fields outside the layer. As the particles leave the layer, they are no longer accelerated and their direction of motion may isotropize quickly in a $B_{\perp}\sim B_0$ external magnetic field, hence effectively reducing the anisotropy of the emission. However, this is not true for the high-energy synchrotron radiation above 100~MeV because the emitting particles have energy above the radiation reaction limit and cool over a timescale shorter than their cyclotron period, {\em i.e.} $t_{\rm syn}(\omega_0/\gamma_{\rm e})\lesssim 1$, similarly if $\beta^{-1}_{\rm rec} (\gamma_{\rm rad}/\gamma_{\rm e})^2\lesssim 1$ (where $\gamma_{\rm rad}$ is given in Eq.~\ref{grad}). As a result, the photons' direction of propagation above 100 MeV should be close to the electrons' direction of motion at the end of the layer. We use this simplifying assumption in Section~\ref{crab}.

\subsection{Impacts from other parameters}

We ran the same simulations for different thicknesses of the reconnection layer ranging from $\delta=3.4\times 10^{10}$~cm to $\delta=3.4\times 10^{14}$~cm~($\approx 0.1 L_{\rm x}$), but we do not find any significant changes in the results. A thick layer tends to accelerate slightly more particles to the highest energy, especially for those that are injected with a high Lorentz factor because their Larmor radius becomes comparable to or smaller than the thickness $\delta$. These particles then have enough time to focus inside the layer where the magnetic field is small. We also investigated the effect of inverse Compton energy losses. The overall shape of the escaping angular distribution does not change, but we note an increase in the density of electrons at higher angle $\theta_{\rm out}$, as expected (see Section~\ref{effect-compton}). The spectrum remains almost monoenergetic but is notably shifted to lower energies if $\mathcal{U}_{\star}\gtrsim 0.1~\mathcal{U}_{\rm B}$, where the maximum Lorentz factor is given by Eq.~(\ref{grad_ic}). We carried out population studies in the presence of magnetic islands in the reconnection layer for $n_{\rm is}=10$,~100 and 1000 islands for different amplitudes $\tilde{B}_{\rm y,max}/B_{\rm y,max}=0.1,~1,~$and 10. We found that the particles cluster quasi-uniformly around the O-points of the islands. The outgoing angular distribution of particles is even more concentrated at small angle $\theta_{\rm out}$ than without islands. The particle energy distribution is also more concentrated at the highest energy $\gamma_{\rm max}=eE_0L_{\rm z}/m_{\rm e}c^2=3.5\times 10^9$ because the particles are channeled in the $z$-direction. These features are more marked for $\tilde{B}_{\rm y,max}/B_{\rm y,max}\gtrsim 1$. Islands increase the anisotropy of the particle distribution and the number of extremely high-energy particles, but the overall distributions are not changed dramatically compared with the case where there are no islands. Note that these results may not be realistic and follow from the assumption of a uniform $E_{\rm z}$. In a highly dynamic plasmoid reconnection regime, $E_{\rm z}$ is expected to be inhomogeneous.

\section{Application: gamma-ray flares in the Crab Nebula}\label{crab}

\subsection{Context and assumptions}

The Crab Nebula is visible throughout the electromagnetic spectrum from radio to very-high energy gamma rays, mostly in the form of non-thermal radiation (see \citealt{2004ApJ...614..897A} for a recent compilation of multiwavelength data). The broad band Spectral Energy Distribution (SED) is essentially composed of a synchrotron bump extending from radio to about 100~MeV energy gamma rays and a second bump attributed mostly to inverse Compton emission emerging above 100~MeV up to about 100~TeV. The overall SED above UV energies is well reproduced by a single radiating population of ultrarelativistic electron-positron pairs injected with a power-law energy distribution immersed in an average $\sim 100$-$200~\mu$G magnetic field ({\em e.g.}, \citealt{2004ESASP.552..439H,2010ApJ...708.1254A,2010A&A...523A...2M}).

The discovery of gamma-ray flares above 100~MeV by the gamma-ray space telescopes {\em AGILE} and {\em Fermi} changed this simple steady-state picture. If the variable emission above 100~MeV is synchrotron radiation, as it is generally believed, and if one associates the duration of the flares ($t_{\rm fl}\sim 1~$day) with the synchrotron cooling timescale of the emitting particles ($t_{\rm s}=3m_{\rm e}c/2r_{\rm e}^2\gamma_{\rm e} B^2_{\perp}$), then the gamma rays are produced by PeV pairs in a milli-Gauss magnetic field, {\em i.e.} in a much stronger field than expected from the modeling of the SED, $B_{\perp}\gg 200\mu$G. This implies that the particles have to be accelerated beyond the radiation reaction limit energy, which is difficult for classical models of particle acceleration unless the emission is substantially Doppler boosted toward the observer \citep{2010MNRAS.405.1809L,2011MNRAS.414.2017K,2011MNRAS.414.2229B,2011ApJ...730L..15Y,2011arXiv1109.1204L}. However, optical and X-ray observations show that the flow is at most mildly relativistic in the nebula ($V_{\rm flow}/c\sim 0.5$, see {\em e.g.} \citealt{2002ApJ...577L..49H}). Alternatively, we have shown in Section~\ref{single_electron} that particles can be efficiently accelerated inside a few light-day-long reconnection layer and a milli-Gauss field to extremely high energies by the large-scale reconnection electric field (Fig.~\ref{fig_spec}). This scenario solves the problem of synchrotron photons above the classical limit $\epsilon_{\rm syn}>\epsilon_{\star}$ \citep{2004PhRvL..92r1101K,2011ApJ...737L..40U}. In addition, the particles angular distribution is expected to be extremely beamed along the current layer, in particular for the most energetic particles emitting above 100 MeV (Fig.~\ref{fig_ang}). This could explain why about $1\%$ of the nebula radiates 5-10 times more flux above 100~MeV than the rest of the system.

Another key feature of the model is the quasi-monoenergetic spectrum of the particles at the end of the reconnection layer (for $B_{\rm z}/B_0<1$, see Fig.~\ref{fig_spec}). This distribution represents the injection spectrum of fresh particles into the nebula and might not always coincide with the spectrum of electrons inferred from the observed gamma-ray emission averaged over the total duration of the flare. In fact, radiative cooling and the (unknown) time-dependent injection of these PeV particles can produce broader distributions like power-laws above 100~MeV as reported by {\em Fermi}\footnote{The spectrum of the September 2010 flare reported by the {\em AGILE} team (averaged over 2~days) is a bit harder than the {\em Fermi} spectrum (averaged over 4~days). This difference could be evidence of synchrotron cooling of the particle distribution \citep{2011ApJ...732L..22V}.} \citep{2011Sci...331..739A}. Reality might be a complex interplay of these two effects. The observation of the detailed spectral evolution during a flare should provide a powerful diagnostic of the injection of PeV particles in the nebula. In addition, even a slight change in the orientation of the reconnection layer with respect to the observer during the flare, for instance due to instabilities in the layer such as tearing and/or kink modes \citep{1998ApJ...493..291B,2011ApJ...728...90M}, can modify the observed emission and explain the short intra-flare variability as well as the symmetric shape of the light-curve profile \citep{2011A&A...527L...4B,2011arXiv1112.1979B}.

\begin{figure}
\epsscale{1.0}
\plotone{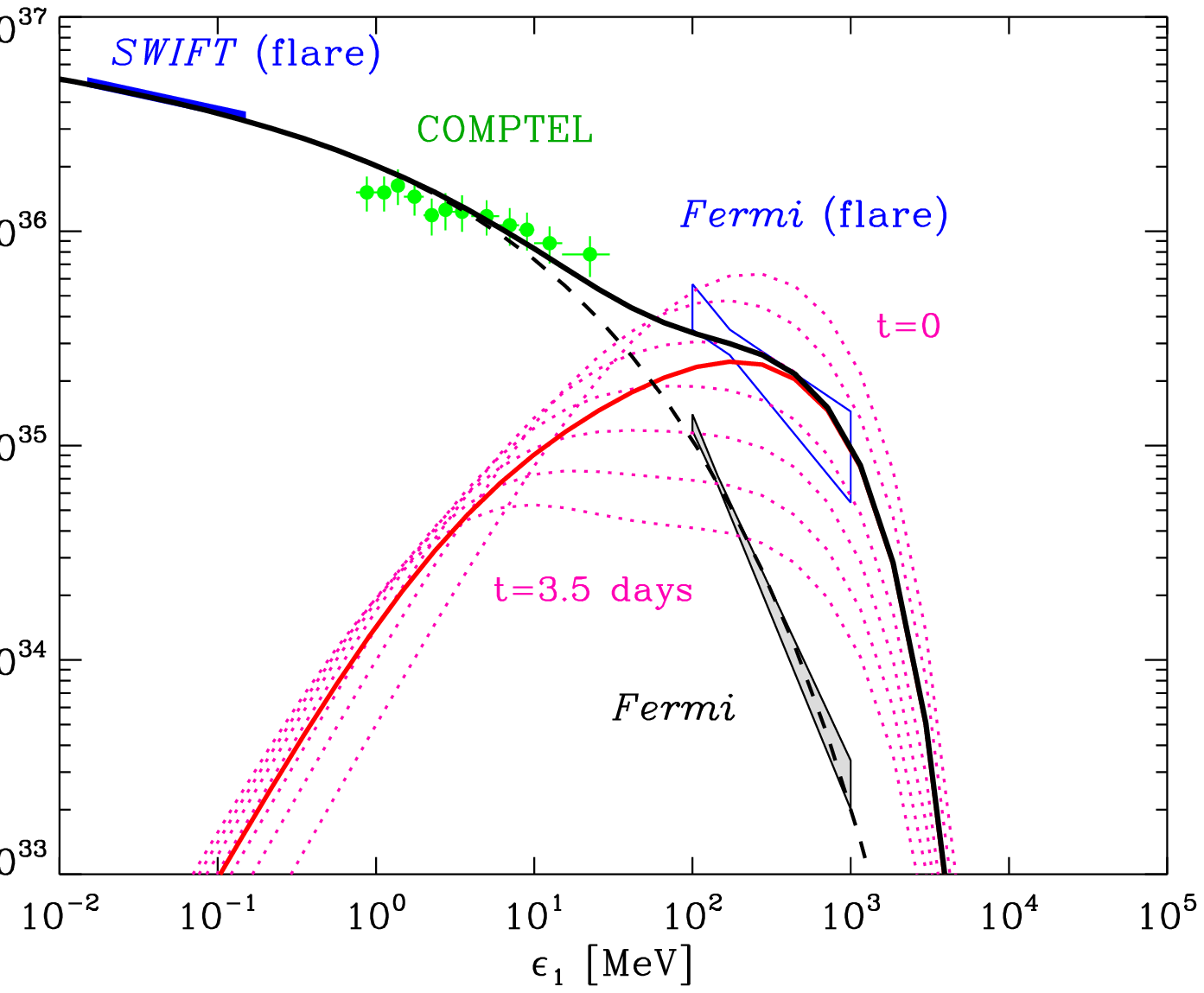}
\plotone{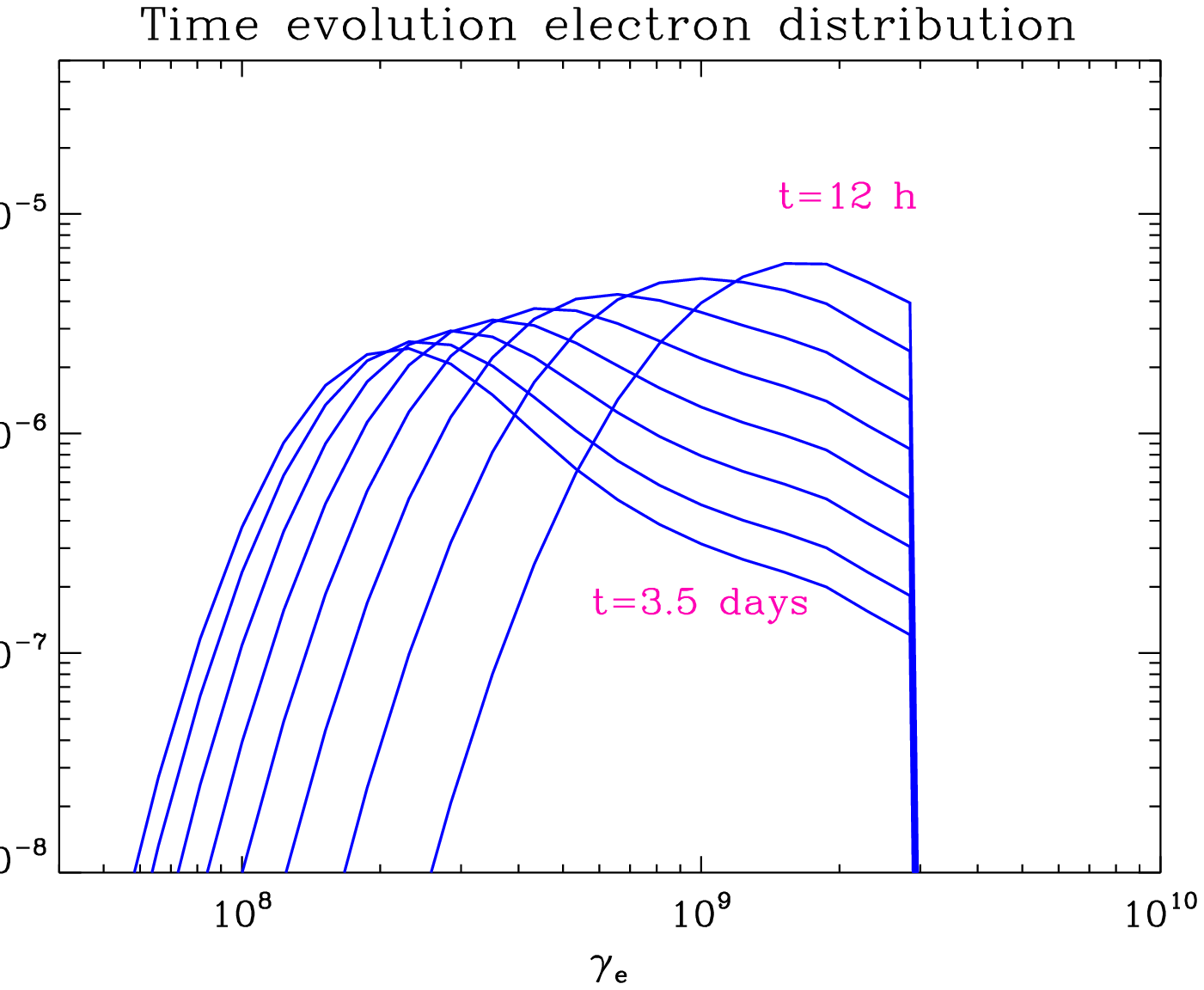}
\caption{{\em Top panel}: Spectral modeling of the Crab Nebula gamma-ray emission during the September 2010 flare. The black dashed line represents the ``quiescent'' spectrum (steady state emission, synchrotron component only) and the red solid line is the synchrotron spectrum of the particles accelerated in the reconnection layer averaged over the duration of the flare (four days). The magenta dotted lines show the evolution of the flaring spectrum sampled every 12~hours (top line $t=0$ and bottom line $t=3.5~$days, see the text for more details about the modeling). The black solid line is the sum of the quiescent and the flaring components. {\em SWIFT} and {\em Fermi} data during the flare \citep{2010ATel.2893....1S,2011Sci...331..739A} are overplotted with blue bowties. Archive data from COMPTEL \citep{2001A&A...378..918K} and the quiescent {\em Fermi} data (only the synchrotron component, \citealt{2011Sci...331..739A}) are plotted for comparison. The distance of the Crab Nebula is fixed at 2~kpc. The flaring inverse Compton emission is negligible. {\em Bottom panel}: Time evolution of the radiating electron distribution sampled every 12~hours injected in the Crab Nebula during the September 2010 gamma-ray flare.}
\label{fig_crab}
\end{figure}

\subsection{Spectral modeling}

For illustrative purposes, we use a simple toy model to reproduce qualitatively the gamma-ray emission observed during the September 2010 flare (Figure~\ref{fig_crab}). First, the ``quiescent'' emission of the nebula is obtained from the synchrotron emission of a cooled and isotropic distribution of electron-positron pairs steadily injected with a power-law energy distribution of index 2.2 for $\gamma_{\rm e}>10^3$, with an exponential cut-off at $\gamma_{\rm cut}=2\times 10^9$, in a 200~$\mu$G magnetic field. On top of the quiescent emission, we start injecting, at the beginning of the flare (marked as $t=0$), another population of particles accelerated in a 4 light-day-long reconnection layer. As discussed in Section~\ref{pop_rad}, it is a good approximation to assume that the $>100$~MeV synchrotron photons are emitted in the same direction as the radiating electron's direction of motion at the end of the layer. Hence, the observed emission during the flare depends strongly on the viewing angle $\theta_{\rm obs}$ of the reconnection layer because of the anisotropy in the outgoing particle distribution (Fig.~\ref{fig_ang}). The emission is maximum if the layer is pointing at the observer ($\theta_{\rm obs}\approx 0$).

For the spectral modeling, we assume that $B_{\rm z}=0$ and $\theta_{\rm obs}\approx 0$. We extract from Figure~\ref{fig_ang} the spectrum of the particles contained in the first bin of the angular distribution ($\theta_{\rm out}=[0,2.5\degr]$) summed over $\phi_{\rm out}$. After this operation, the spectrum of the particles is composed only of the most energetic particles with $10^9<\gamma_{\rm e}\lesssim 3\times 10^9$ (See Fig.~\ref{fig_crab}, \emph{bottom} panel). Once injected in the nebula, particles are no longer accelerated and are subject only to synchrotron cooling in a $B_0=5~$mG disordered magnetic field. We arbitrarily choose that the number of fresh particles supplied per unit of time by the layer decreases exponentially with time ${\rm d}\dot{N}_{\rm inj}/{\rm d\gamma_{\rm e}}={\rm d}\dot{N}_0/{\rm d}\gamma_{\rm e}\exp(-t/\tau) $, with a decay timescale $\tau=1~$day. The particle spectrum time evolution is obtained by solving the Fokker-Planck equation using the implicit Chang-Cooper algorithm \citep{1970JCoPh...6....1C} and is shown together with the resulting synchrotron spectrum time evolution in Figure~\ref{fig_crab}. The electron and gamma-ray spectra soften and broaden slightly with time. Nonetheless, little flux is expected below 100~MeV compared with the emission from the rest of the nebula. Averaged over the duration of the flare, the computed spectrum reproduces qualitatively the observations (Fig.~\ref{fig_crab}, \emph{top} panel). We also computed the inverse Compton emission associated with the flare. Following \citet{2010A&A...523A...2M}, the target photon density is composed of the synchrotron photons from the nebula, thermal emission at 93~K from the dust and the 2.7~K cosmic microwave background. We find that the expected inverse Compton emission is far too weak (by several orders of magnitude) compared to the quiescent Compton emission to increase the very-high energy flux noticeably, because the energy density in the background radiation field is much smaller than the magnetic energy density in the Crab Nebula. The inverse Compton drag force has a negligible effect on the motion of the particles. In this model, the gamma-ray flares are confined to the high-energy band $>100~$MeV up to a few GeV and do not have any detectable counterparts at other wavelengths.

\subsection{Energetics of the flare}

The total number of pairs required in the spectral modeling to account for the September gamma-ray flare luminosity above $100$~MeV is $N^{\rm iso}_{\rm inj}=8.5\times 10^{37}$ pairs (if the Crab Nebula is located at 2~kpc). This number is not corrected for the extreme anisotropy of the radiating particle distribution. Taking the beaming effect into account, the actual number of pairs needed is $N_{\rm inj}=N^{\rm iso}_{\rm inj}(\Omega/4\pi)\approx 7\times 10^{35}~$pairs (with $\Omega\approx 0.1$, see Section~\ref{pop}). The mean energy of the particles injected by the layer into the nebula is $\langle\gamma_{\rm e}\rangle\approx 2.3\times 10^9$. Then, the total energy required in these PeV particles is about $\mathcal{E}_{\rm Flare}\approx 10^{39}~$erg. This energy represents only about 0.2\% (about 35\% if there is no beaming) of the total magnetic energy available in the flare region given by
\begin{eqnarray}
\mathcal{E}_{\rm mag} & = & \frac{c}{4\pi}E_{\rm z}B_0\times 2\left(2L_{\rm x} L_{\rm z}\right)t_{\rm Flare} \nonumber \\
& = & 4\beta_{\rm rec}\left(\frac{B_0^2}{8\pi}\right)\left(ct_{\rm Flare}\right)^3\approx 4.4\times 10^{41}~\rm{erg},
\end{eqnarray}
{\em i.e.}, the Poynting flux passing through the surface of the layer $(2L_{\rm x}L_{\rm z})=(ct_{\rm Flare})^2$ in the $\pm y$-direction times the duration of the flare.

We conclude that this scenario provides a plausible solution to the Crab gamma-ray flares paradox.

\subsection{Possible locations of the reconnection layer}

Our model does not specify the location of the reconnection layer responsible for the gamma-ray flares, but one can envisage at least three plausible sites: within the pulsar wind upstream of the wind termination shock, just outside the termination shock in the equatorial region, or outside the termination shock in the polar regions. In the case of a misaligned rotator, the magnetic field in the wind is expected to have a toroidal striped-like structure of alternating polarity around the equatorial plane (the so-called ``striped wind'', see \citealt{2009ASSL..357..421K} for a review). Magnetic reconnection could occur in this configuration and convert the magnetic energy into particle kinetic energy \citep{1990ApJ...349..538C,1994ApJ...431..397M}. The length of the layer would be of order the distance to the pulsar. However, \citet{2001ApJ...547..437L} found that the wind in the Crab should reach the termination shock before a substantial fraction of the magnetic energy is dissipated, unless the wind contains a high density of pairs \citep{2003ApJ...591..366K}.

If the wind remains highly magnetized up to the termination shock, the striped structure could be rapidly dissipated by shock-driven magnetic reconnection in the equatorial regions downstream of the shock \citep{2003MNRAS.345..153L,2007A&A...473..683P,2008ApJ...682.1436L,2011ApJ...741...39S}, similarly to what may happen in the nonrelativistic case of the solar system close to the heliospheric termination shock \citep{2010ApJ...709..963D}. The current layer could be of order the termination shock radius, $\sim 0.1~$pc in the Crab, which is too long to explain the short variability timescale of the flares, but fragmentation of the current sheet could lead to more rapid variability. The amount of magnetic energy available for dissipation in this case would be limited by the pulsar misalignment angle, with a high efficiency possible only if the angle is close to $90^\circ$.

The third possibility is that rapid reconnection is triggered by current-driven MHD instabilities downstream of the shock. If the magnetic field is dynamically important, the postshock wind will try to relax to a configuration in which gas pressure forces balance the stresses due to the predominently toroidal magnetic field \citep{1992ApJ...397..187B}. Such a system is highly unstable to kink instabilities (and slightly less unstable to pinch instabilities) in regions where the plasma beta parameter is $\ga O(1)$, on time scales of order the Alfv\'en crossing time \citep{1998ApJ...493..291B,2011ApJ...728...90M}. The nonlinear development of the instability leads to a complex system of current sheets (O'Neill et al., in preparation) that could rapidly dissipate most of the magnetic energy at mid- to polar latitudes. The dissipation of the magnetic energy in the polar region could power the X-ray jet in the Crab. New observations with Chandra show that the southern jet has moved since 2000 \citep{2011arXiv1111.3315W}, perhaps related to the kink instability. In addition to the formation of current layers in the polar regions, it is expected that the magnetic field strength is amplified close to the rotational axis by the z-pinch, as required to power the flares.

In addition to setting up the conditions for extreme particle acceleration, the establishment of reconnection sites due to current-driven instabilities could resolve the long-standing ``sigma problem''. Hot gas in polar regions where the magnetic field has dissipated could exhaust to larger radii, allowing more highly magnetized gas from equatorial latitudes to fill the void. As it does so, this fresh gas will also go unstable, leading to the destruction of its magnetic energy, and so forth. Thus, current-driven instabilities at high latitudes could set up a conveyor belt leading to the dissipation of most of the magnetic energy introduced by the shocked pulsar wind. Such a process could reconcile the relatively high magnetic field (and inferred wind $\sigma \sim O(1)$) required by the particle acceleration model with the low $\sigma$ required by models of the nebula \citep{1974MNRAS.167....1R,1992ApJ...397..187B}.

\section{Conclusion}\label{ccl}

The existence of synchrotron photons above 100~MeV emitted by $e^+/e^-$ pairs requires the non-standard condition that the electric field exceeds the magnetic field in the acceleration region. This condition can be fulfilled in regions where ideal MHD approximation breaks down, such as reconnection sites.

Relativistic test-particle simulations presented in this paper show that high-energy particles moving across a thin reconnection layer are accelerated by the electric field and focus rapidly deep into the layer where $E>B$. This mechanism transforms any initial distribution of particles into an extremely beamed distribution in the reconnection layer. In addition, our population studies show that the energy distribution of the particles at the end of the layer piles up to the maximum potential drop energy available, $eEL_{\rm z}$, suggesting that the reconnection layer acts almost as a pure linear accelerator. Because the perpendicular magnetic field is small deep inside the layer, individual particles can exceed the nominal radiation reaction-limited energy and emit $>100$~MeV synchrotron photons provided that the layer is long enough. We note that a strong guide field and the associated in-plane electric field tend to defocus the particles outside the layer, decreasing the fraction of particles accelerated to the highest energies. We find also that static magnetic islands in the layer, generated for instance by tearing instabilities, tends to trap and focus the particles toward the O-points. If the layer is bent by {\em i.e.} kink instabilities, particles' trajectories follow the curvature of the layer. The final energy of the particles and the general focusing mechanism described in this article remain unchanged by these deformations as long as the curvature length-scale is not too short compared with the distance traveled by the particles between two crossings through the layer.

These key features of the model ({\em i.e.} generation of a fan beam of particles with energy $\sim eEL_{\rm z}$) are essential to explain the apparent high number of PeV electrons required to power the Crab Nebula gamma-ray flares. For illustrative purposes, we applied the model to the September 2010 flare. Observations can be qualitatively reproduced assuming a few light-day-long reconnection layer and a few milli-Gauss reconnecting magnetic field. The resulting spectrum is close to the synchrotron spectrum of a single electron. Cooling and time-dependent injection of fresh particles in the nebula during the flare can broaden and soften the spectrum averaged over the duration of the flare. The strong beaming of the electron distribution emitting the high-energy radiation accounts for the overall luminosity of the flare, if the reconnection layer points toward the observer. This implies that gamma-ray flares could happen frequently in the nebula, but only those occasionally pointing at us are detected. The inverse Compton emission associated with the flare is far too low to be observed and no detectable counterparts of the flare at other wavelengths are expected in this model.

The mechanism of particle acceleration described in this article may be at work in other astrophysical objects, for instance in active galactic nuclei (AGN) jets (Nalewajko et al., in preparation). A few blazars flaring in TeV gamma rays exhibit variability timescales ($\sim$ minutes) much shorter than the light-crossing time of the Schwarzschild radius of the central supermassive black hole ($\sim$ hour) \citep{2007ApJ...664L..71A,2007ApJ...669..862A,2011ApJ...730L...8A}, suggesting that particles are quickly and efficiently accelerated in a small volume of the relativistic jet. The acceleration and, more importantly, the focusing of the particles within a reconnection layer could account for strong beaming of the TeV flare emission. In the environment of AGN jets, the effect of the inverse Compton drag force becomes important and would decrease the maximum energy reached by the particles at the end of the layer. Nevertheless, we found that the distribution of particles should remain extremely anisotropic. Note that this model differs from the ``mini-jet'' model \citep{2009MNRAS.395L..29G,2010MNRAS.402.1649G,2011MNRAS.413..333N} in which particles are assumed to be beamed just by the relativistic bulk motion of the reconnection outflow (moving perpendicular to the electric field, in the jet comoving frame) expected in a Poynting flux-dominated jet ($\Gamma_{\rm bulk}\sim \sigma^{1/2}$, where $\sigma\gg 1$ is the ratio of the Poynting flux to the particle kinetic energy flux). In the Crab Nebula, mini-jets are unlikely to form because $\sigma$ is at most of order unity. 

The test-particle simulations presented in this work are not fully satisfactory, because many uncertainties remain regarding the structure of the electromagnetic fields inside and outside the reconnection layer. Interactions between particles are also neglected in this study. It is unclear what the effect of extremely high-energy particles on the dynamics of the reconnection layer would be. A self-consistent and more realistic approach to this problem would be to use PIC simulations with ultrarelativistic electron-positron pairs, including the radiation reaction force as pioneered by \citet{2009PhRvL.103g5002J}. Another challenge is to calculate accurately the radiation emitted by the particles during the acceleration process inside and outside the reconnection layer (time dependent angular and energy distributions). These issues are left for future investigations.

\acknowledgements The authors thank J.~G.~Kirk for discussions about the radiation reaction force, R.~D.~Blandford and R.~Buehler for valuable discussions regarding the Crab gamma-ray flares, K.~Nalewajko, and M.~Yamada and the referee for their comments. This work was supported by NSF grant PHY-0903851, NSF grant AST-0907872 and NASA Astrophysics Theory Program grant NNX09AG02G.

\bibliography{speiser}

\end{document}